\documentclass[onecollarge]{svjour2} 

\usepackage{amsmath,amssymb}
\usepackage{color,ulem}
\usepackage{graphicx}

\def\cO {{\cal O}} 

\begin{document}
\title{Fractal dimensions of self-avoiding walks and Ising high-temperature graphs in 3D conformal bootstrap}
\author{Hirohiko Shimada         \and
Shinobu Hikami 
}
\institute{Mathematical and Theoretical Physics Unit, Okinawa Institute of Science and Technology Graduate University,  Onna, Okinawa, 904-0495, Japan\\
\email{hirohiko.shimada@oist.jp}\\
\email{hikami@oist.jp}
}

\maketitle
\begin{abstract}
The fractal dimensions of polymer chains and high-temperature graphs in the Ising model both in three dimension are 
determined using the conformal bootstrap applied for the continuation of the $O(N)$ models from 
$N=1$ (Ising model) to $N=0$ (polymer). 
The unitarity bound below $N=1$ of the scaling dimension for the the $O(N)$-symmetric-tensor
develops a kink as a function of the fundamental field as in the case of the energy operator dimension in the Ising model. 
Although this kink structure becomes less pronounced as $N$ tends to zero, 
an emerging asymmetric minimum in the current central charge $C_J$ can be used to locate the CFT.  
It is pointed out that certain level degeneracies at the $O(N)$ CFT should induce these singular shapes of the unitarity bounds.
As an application to the quantum and classical spin systems, we also predict critical exponents 
associated with the $\mathcal{N}=1$ supersymmetry,  
which could be relevant for locating the correspoinding fixed point in the phase diagram.
\end{abstract}

\section{Introduction}

Conformal field theory (CFT) is an indispensable framework in deepening our understanding on the universality class of the critical phenomena which goes beyond the renormalization group (RG).
Despite its incomparable success in 2D, the clues for 3D CFT has been scarce until recently. 
The recent breakthrough came from numerical determination for the 3D Ising exponents \cite{El-Showk,El-Showk14} using the crossing-symmetry sum rule for the $\mathbb{Z}_2$-symmetric intermediate states
in the four-point function $\langle \phi\phi\phi\phi \rangle$ of the same scalar field (the fundamental field $\phi$ in the $\lambda \phi^4$-theory). 
The key empirical observation, which becomes a cornerstone in this so-called conformal bootstrap approach, was that the scaling dimensions
of the spin and energy operator in the Ising model corresponds to      
a ``kink" that emerges along the unitarity upper-bound curve for the dimension $\Delta_{\phi^2}$ of the leading non-trivial 
$\mathbb{Z}_2$ symmetric operator $\varepsilon=:\phi^2:$ as a function of the dimension $\Delta_{\phi}$ of the fundamental field. 
This singular shape, the kink in the unitarity bound, is shared also in the case of
the sum rule for the $O(N)$-symmetry, which can be used to map
the critical $O(N)$ model 
with $N=2,3,\cdots, \infty$ on the $\Delta_{\phi}$-$\Delta_S$ plane \cite{Kos14} 
, where $\Delta_\phi$ and $\Delta_S$ respectively stand for 
the dimension of the fundamental field $\phi_a$ (``$a$" is an $O(N)$ label) 
and the dimension of the energy operator  $\varepsilon=\sum_{a} :(\phi_a)^2:$, which is 
the leading non-trivial operator in the $O(N)$ singlet sector $S$. 
 
Towards an analytic understanding on the consequence of the 3D conformal symmetry, it would be important to 
aim at a representation theory of the spectrum generating algebra analogous to the degenerate representation in the Virasoro algebra \cite{Belavin}.
Another outstanding direction would be to generalize the ideas in the stochastic Loewner evolution (SLE) \cite{Bauer} so as to describe {\it critical geometry} in 3D. 
In this respect, the importance of the continuous family of the critical $O(N)$ models below $N=2$ could be emphasized more 
since in 2D they precisely represent the continuous family of the models 
described by the SLE$_\kappa$ with $2 \leqslant \kappa\leqslant 4$ via the trigonometric relation 
\begin{align}\label{kappa}
\kappa=4\pi/\arccos(-N/2).
\end{align}
Mathematicians proved that the Hausdorff dimension of SLE$_\kappa$ curves is given by 
\begin{align}\label{Beffara}
d_F(\kappa)=1+\kappa/8,
\end{align}
(Beffara's theorem \cite{Beffara}). In physics, the same fractal dimension can be computed from the dimension of the 2-leg operator (a special case of the watermelon operator for an arbitrary number of legs
\cite{Duplantier}) 
represented by an $O(N)$ symmetric tensor operator $\varphi_{ab}$,  which behaves as a scalar under the spatial $O(2)$ rotations.

In this paper, we study the 3D $O(N)$ model below $N=2$ with a focus on the fractal dimension of the loops in the high-temperature expansion.
The fractal dimension may be given by $d_F = D - \Delta_T$, where $\Delta_T$ is the scaling dimension 
\footnote{This dimension of the relevant operator in $T$-sector is denoted by $\Delta_T$ following the convention in \cite{Kos14}.
This ``$T$" should not be confused with the stress-energy tensor $T^{\mu\nu}$, which has spin $2$ with $O(D)$ and the fixed scaling dimension $D$.} 
of the most relevant operator $\varphi_{ab}$ in the $O(N)$ symmetric tensor sector $T$. 
Apart from the models with $N\geqslant 2$, where $\Delta_T$ has been estimated \cite{Kos14}, 
there are several important physical cases in 3D, where   
understanding based on the conformal symmetry,
in particular, the determination of the fractal dimensions may be interesting. 

\vspace{2mm}
\underline{(a) Polymer ($N=0$)} The $N\to 0$ limit of the $O(N)$ symmetry, where the degeneracy of $\Delta_T$ and $\Delta_S$ occurs, 
describes polymer chains under the excluded volume limit (a self avoiding walk)
as shown in the celebrated work by de Gennes \cite{deGennes}.
A direct approach to this polymer limit makes various OPE coefficients singular and makes 
the current bootstrap method, which hinges on the positivity of the squared OPE coefficients (a main part of the unitarity), difficult.
For instance,  the square of the OPE coefficient  $\lambda^{\phi\phi}_{T}$ for the stress-energy tensor may have a simple pole at $N=0$ since the Ward identity tells us that it is inverse proportional to the central charge $C_T$, which is essentially proportional to the number of components $N$.
Practically, as $N$ tends to zero, this pole
seems to result in an effective slowdown of the convergence to the optimal unitarity bound,
meaning that the number of derivatives necessary to attain a given precision increases more rapidly.
Accordingly, the detection of the kink at $N=0.1$ within a limited computational cost becomes much more difficult compared with the case of finite $N$ (e.g. $N=2$).
We circumvent this difficulty by assuming that a clear change of the slope  $\partial C_J/\partial\Delta_\phi$ 
for current central charge $C_J$ defined through the 
conserved current $J^\mu_{ab}$ \cite{Petkou96} may correspond to the dimension $\Delta_\phi$ of the CFT.
This analysis leads to the estimate for the fractal dimensions $d_F=3-\Delta_T(0)\sim 1.701$.

\vspace{2mm}
\underline{(b) The Ising model ($N=1$) and the $\mathcal{N}=1$ SUSY fixed point}
The operator content of the $O(N)$ model at $N=1$ 
contains that of the Ising model as its singlet sub-sector   
and its thermodynamical exponents can be determined from the well-studied dimensions
$\Delta_\phi$ for the spin operator and  $\Delta_{\phi^2}=\Delta_S$ for the energy operator.
Since the $O(1)$ model contains only one component scalar, it is less noticeable that the dimension $\Delta_T$ of the symmetric ``tensor" $\varphi_{ab}$ may carry important information on the critical exponents. 
It is, however, natural to consider that $\Delta_T$  is one of the geometric exponents that determines 
the fractal dimension $d_F=3-\Delta_T(1)\sim 1.734$ for the high-temperture graphs 
also measured by a Monte Carlo (MC) simulation for the 3D Ising model \cite{Winter}. 
As a natural extension of this analysis, 
we also give the fractal dimension which would possibly be realized by the magnetic flux loops in an effective gauge theory,
which appears, for instance, 
in the Kitaev model plus 
the local exchange interaction of the Ising type \cite{Kamiya}.
An interesting possibility is that the phase diagram of this model, ``magnetic three-state of matter" 
(or its slight extension), may contain the fixed point of the 3D $\mathcal{N}=1$ superconformal field theory (SCFT), 
whose 2D counterpart is a well-established SCFT \cite{Friedan}, which may explain 
the Majorana fermion nature of the 2D Ising model
as the Nambu-Goldstone fermions associated with a spontaneous breaking of the supersymmetry. 
Unlike in 2D, where the $\mathcal{N}=1$ SCFT corresponds to the universality class of the Ising tricritical point \cite{Friedan}, 
our view is that the SCFT and the Ising tricritical point are distinct fixed points in 3D. 
This 3D SCFT is also proposed as a boundary effective theory for the topological superconductor \cite{Grover}.


\vspace{2mm}
\underline{(c) The model at $N=-2$ and its possible relation to the loop-erased random walk} 
The $O(N)$ model at $N=-2$ may be considered as an endpoint of the continuous family of the $O(N)$ model
in the sense that the dimension for the fundamental field and energy operator reduce to the mean field values
\footnote{
The non-renormalization of these dimensions is due to the topological property of the $O(N)$-vertex:
$\sum_{d} \delta_{cd}\cdot \left(\delta_{ab}\delta_{cd} + \delta_{ac}\delta_{bd} + \delta_{ad}\delta_{bc} \right)=(N+2)\delta_{ab}$,
which vanishes at $N=-2$, regardless of the space dimension $D$ \cite{Balian}.}
$(\Delta_{\phi}, \Delta_S)$=$(1/2,1)$ 
and may be paired with the other end point $N=\infty$, 
where the mean field value of $(\Delta_{\phi}, \Delta_T)=(1/2,1)$ and the spherical model value of $\Delta_S=2$ are realized.
Among these operators in $N=-2$ and $N=\infty$, the only nontrivial dimension is $\Delta_T(-2)$; as in 2D \cite{Lawler},
it would be natural to conjecture that  $d_F= 3-\Delta_T(-2)\sim 1.614$ is the fractal dimension of the loop-erased random walk.

Apart from the MC simluations already mentioned, there are still vast works of related simulations, among which some notable are  
certain sophisticated tests of the conformal invariance in the 3D self-avoiding walk ($N\to 0$) \cite{Kennedy},
the worm algorithm that can be applied for continuous values of $N\geqslant 0$ \cite{Liu},
and certain clever algorithms with analysis that get rid of the correction-to-scaling to attain 
ever improving precision on the self-avoiding \cite{Clisby} and loop-erased \cite{Wilson} random walks.

Our emphasis is not on the precision for the critical exponents, though some of them including perhaps 
the anomalous dimensions slightly above $N=0$ and the fractal dimension for the Ising model may already be more accurate than existing MC simulations \cite{Liu,Winter}.  
Instead, it is our purpose here to consider how the conformal invariance may be used to determine the fractal dimension,
without any use of machine generated random numbers, and
to help opening a way to understand more theoretical aspects (such as the kink formation, the representation theory, the 3D SLE, and so on) in the 3D $O(N)$ CFT in general.

This paper is organized as follows.
We consider the $O(N)$ model for a global range of $-2\leqslant N\leqslant \infty$ in Section \ref{section:ONglobal},
and show that the fractal dimension $d_F$ can be regarded as a geometric RG eigenvalue given by the dimension $\Delta_T$ of the traceless symmetric tensor $\varphi_{ab}$.
Section \ref{section:degeneracy} is a quantitative discussion on how the gap between $\Delta_T$ and $\Delta_S$ closes in the polymer limit $N\to 0$
using a simple 6-loop RG analysis leaving the details in Appendix.
In Section \ref{section:sumrule}, the intermediate states in the four point function is classified into three sectors (S: singlet, T: traceless symmetric tensor, A: antisymmetric tensor)
using the operator product expansion (OPE) $\phi_a \times \phi_b$ of the fundamental fields.
The key equation in the $O(N)$ conformal bootstrap, namely, the crossing symmetry sum rule is reviewed 
with a brief discussion of the solution manifold with regard to the unitarity bound.
In Section \ref{section:cTcJ}, the definitions and useful $1/N$-expansions of the current central charge $C_J$ as well as those of the standard central charge $C_T$ are given. 
The implication of the unitarity and corresponding implementation of the bootstrap, though being standard, are given in Section \ref{section:bootstrap}.
We present our main results in Section \ref{section:singularshape}.
We give a qualitative description on 
how an effective smoothing of the kink in $\Delta_T$ occurs in the polymer limit $N\to 0$ 
(we idenfify it as {\it a severe unitarity wall}, across which the continuation of the unitarity-saturating solution is interrupted) 
and
discussions on how certain level-degeneracies in the $O(N)$ CFT would be related to
various singular shapes (the kinks in $\Delta_T$, $C_T$, and in particular, $C_J$) 
of the unitarity bounds
in Section \ref{section:kink} and Section \ref{section:slopechange}, respectively. 
We determine the fractal dimensions by the conformal bootstrap for the polymers ($N\to 0$) in Section \ref{section:N=0} and 
for the 3D Ising high-temperature graphs ($N=1$) in Section \ref{section:N=1}.
We compute the fractal dimension for $N=-2$ in RG and conjecture that it corresponds to that of the loop erased random walk in Section \ref{section:N=-2}.
In Section \ref{section:SUSY},  we estimate the set of the scaling dimensions $(\Delta_{\phi}, \Delta_{\phi^2})$ for the $\mathcal{N}=1$ SCFT 
and discuss the relation to the critical exponent $\nu$ as well as the fractal dimension $d_F|_{SUSY}$ for the corresponding excitation.
We conclude with selected future directions in Section \ref{section:conclusion}.


\section{The $O(N)$ CFT for a global range of $-2\leqslant N\leqslant \infty$}


\subsection{The fractal dimension and the traceless symmetric tensor $\phi_{ab}$}\label{section:ONglobal}
We start with the discussion on the two relevant operators $\varepsilon$ and $\varphi_{ab}$ in the $O(N)$ model, which respectively belong to
the $O(N)$-singlet sector ($S$) and the $O(N)$-symmetric tensor sector ($T$). These operators are formed as a bilinear of the fundamental field 
$\phi_a$ with the scaling dimension $\Delta_\phi$, which transforms as a fundamental representation of the $O(N)$ group:
\begin{align}
S:& &\varepsilon(x) &= \sum_{a=1}^N :\phi_a^2:, &\\
\label{phi_ab}
T:& &\varphi_{ab}(x) &= :\phi_a\phi_b: - \frac{\delta_{ab}}{N} \sum_{c=1}^N :\phi_c^2:. &
\end{align}
The energy operator $\varepsilon$ is already in the single component model  and plays an essential role 
in the initial formulation of the conformal bootstrap for the Ising model, which has 
the $\mathbb{Z}_2=O(1)$ symmetry \cite{El-Showk}. 
The most relevant operator $\varphi_{ab}$ in $T$ sector is responsible for the crossover phenomena
with respect to the symmetry breaking $O(N) \to O(M) \times O(N-M)$ with an arbitrary $M$.
As in the general statistical model, the two-point function $\left\langle \phi_a(x) \phi_b(y)\right\rangle$   
can be expressed as a sum over self-interacting random walks between $x$ and $y$ \cite{Itzykson} (also \cite{Parisi,Feynman}). 
The Hausdorff dimension of this random walk is given by $d_F=\phi_2/\nu$ \cite{Kiskis}, where $\nu$ and $\phi_2$  are 
respectively the correlation length exponent and the crossover exponent \cite{Hikami} of the $O(N)$ model.
Since these two independent exponents are related to the scaling dimensions $\Delta_S$ of $\varepsilon$ and 
$\Delta_T$ of $\varphi_{ab}$ by
\begin{align}
\nu=\frac{1}{ D- \Delta_S}, \qquad \phi_2= \frac{D-\Delta_T}{D-\Delta_S},
\end{align}
one has a simpler expression for the Hausdorff dimension
\begin{align}\label{d_F}
d_F = D-\Delta_T,
\end{align}
which may be viewed as a geometric RG eigenvalue $y_G$ in the light of the fact \cite{CardyScaling} 
that the magnetic (thermal) RG eigenvalues can be determined by the relation $y_H=D-\Delta_\phi$ ($y_T=D-\Delta_S$).

In the range $-2\leqslant N \leqslant \infty$, this dimension $\Delta_T$ decreases monotonically from a certain value $\Delta_T(-2)$ 
(FIG. \ref{fig:NDelta} and below for the meaning)
to the trivial value $\Delta_T(\infty)=D-2$ 
and at $N=0$ crosses the dimension $\Delta_S$ of the energy operator $\varepsilon$ (in the $O(N)$ singlet sector $S$, as mentioned), 
which in turn increases monotonically from $\Delta_S(-2)=D-2$ to $\Delta_S(\infty)=2$ in the same range of $N$
(one may also notice the asymptotic slopes computed in $1/N$-expansions are symmetric in 3D: $\frac{\partial}{\partial\Delta_\phi}\Delta_S(\infty)=-\frac{\partial}{\partial\Delta_\phi}\Delta_T(\infty)=8$).
Actually, this somewhat dual behavior of $\Delta_T$ and $\Delta_S$ in the global range of $N$ is almost shared in the 2D $O(N)$ model though the range $N\in [-2, \infty]$ 
should be replaced
by $N\in [-2,2]$, where the model has a critical point and exact results are available from the Coulomb gas \cite{Nienhuis}, SLE \cite{Bauer}, and CFT torus partition funtion (as described just below) for continuous values of $N$; it is also likely to be a generic feature of the $O(N)$ CFT in $2\leqslant D<4$ from the RG point of view. 

The operator content of the 2D $O(N)$ model with $-2\leqslant N\leqslant 2$ can be studied exactly by the torus partition function \cite{diFrancesco}.
Using the Coulomb gas coupling $g$ ($1\leqslant g\leqslant 2$) determined by the relation $N=-2\cos(\pi g)$,
the central charge (which is a 2D counterpart of $C_T$ in Section \ref{section:cTcJ}) and the scaling dimensions are given by
\begin{align}
c=1-\frac{6(g-1)^2}{g},\quad \Delta_\phi=1-g/2-3/(8g),\quad \Delta_S=4/g-2,\quad \Delta_T=1-1/(2g).
\label{2D}
\end{align}
Since the SLE parameter $\kappa$ is actually related to $g$ by $\kappa=4/g$, the last relation in \eqref{2D} with \eqref{d_F}  is equivalent to
 the formula \eqref{Beffara} in the Beffara's theorem \cite{Beffara}. 
 
In the 2D torus partition function, the multiplicity $N(N+1)/2-1$$=$$(N-1)(N+2)/2$ for the traceless symmetric tensor $\varphi_{ab}$ tends to zero 
as $N\to1$ in accordance with the observation that
the expression in \eqref{phi_ab} apparently vanishes at $N=1$. 
It is, however, instructive to note that the dimension $\Delta_T=5/8$ in the $N=1$ model ($g=4/3$) is of physical relevance.
Namely, it corresponds via \eqref{d_F} to the fractal dimension $d_F=11/8$ of the Ising interfaces, which are the SLE$_{\kappa=3}$ curves.  
The relevance of this tensor $\varphi_{ab}$ for generic $D\geqslant 2$ in the $O(N)$ sum rule at $N=1$ will be discussed in the end of Section \ref{section:sumrule} and will be used to determine its scaling dimension in $D=3$ in Section \ref{section:N=1}.
Similarly for the 2D $N=-2$ model, 
the dimension $\Delta_T=3/4$  ($g=2$) leads to the fractal dimension $d_F=5/4$ of the loop-erased random walks (SLE$_{\kappa=2}$ curves) \cite{Lawler}.

We will give in Section \ref{section:N=-2} a simple estimate for $d_F$ in the $N=-2$ model using \eqref{d_F} by a pseudo $\epsilon$ expansion in the 6-loop RG, which agrees with the numerical simulations results obtained for the 3D loop-erased random walk 
\cite{Guttmann,Agrawal,Grassberger,Wilson}. 
We use the conformal bootstrap to determine 
$\Delta_T$ in the $O(1)$ model and the fractal dimension $d_F$ of the high-temperture graphs 
in the 3D Ising model \cite{Winter} in Section \ref{section:N=1}.
\begin{figure}
\includegraphics[width=16cm]{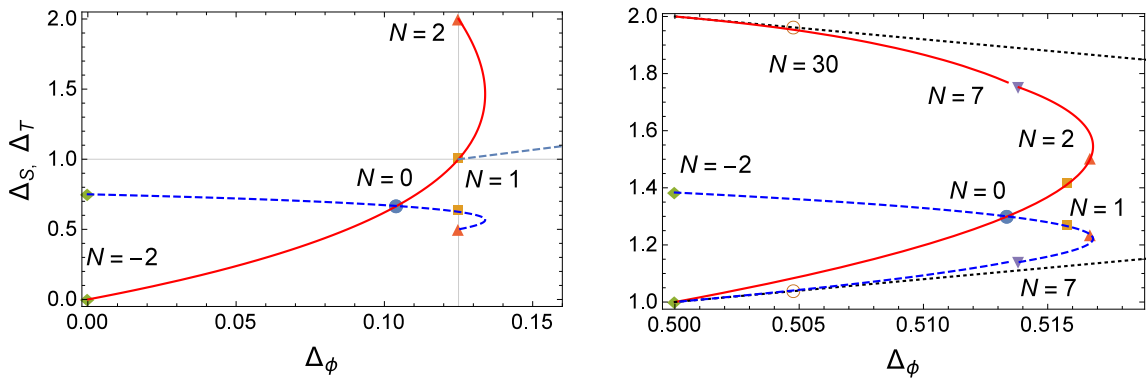}   
	\caption{ \footnotesize
The scaling dimensions of the singlet scalar $\varepsilon$ ($\Delta_S$: solid red) and traceless symmetric tensor $\varphi_{ab}$ ($\Delta_T$: dashed blue) as a function of $\Delta_\phi$ 
in 2D (left: eq.\eqref{2D}) and in 3D (right). The right branch of the unitariy bound (dashed gray) with the $\mathbb{Z}_2$ case \cite{Rychkov} is shown for 2D as a guide to the eye. 
The 3D curve is obtained as the $[5/1]$-Pad\'{e} approximant for $N\in [-2,7]$ continued by the curve from the pseudo $\epsilon$-series (Appendix) for $N>7$ and should be 
regarded as schematic as the anomalous dimension tends to be smaller than the genuine value.
The $N\to \infty$ asymptotics (dotted) are shown for both scaling dimensions.
\label{fig:universalcurves}          } 
\end{figure}

\subsection{The degeneracy of the relevant operators from $S$ and $T$ sectors in the limit $N\to 0$}\label{section:degeneracy}

In addition to the above two important cases, we are especially interested in the $N\to 0$ limit of the 3D $O(N)$ model, 
which describes dilute solutions of polymers, where the random walk becomes self-avoiding. 
Besides such physical relevance, the limit $N\to0$ is theoretically special for the following two reasons.
First, some of squared OPE coefficients may become negative for $N<0$ due to single poles at $N=0$, 
which makes it difficult to take the approaches based on the unitarity, on which most of the present conformal bootstrap schemes depend.       
Second, as mentined above, $N=0$ is the precisely the point where the degeneracy of the two scaling dimensions $\Delta_S$ and $\Delta_T$ take place.
As a quick example using \eqref{2D} in 2D, $\Delta_S=\Delta_T=2/3$ ($d_F=4/3$) 
\footnote{It is also well known in 2D that $d_F=4/3$ coincides with the Flory value \cite{Flory} $d_F=\nu^{-1}\sim(D+2)/3$.}
 follows from $g=3/2$ for $N=0$ and the gap opens with the following asymmetric $N$-derivatives,
\begin{align}
\left.\frac{\partial}{\partial N}\right|_{N=0}\hspace{-5mm}\Delta_S=8/(9\pi)=   0.282942\cdots     ,\qquad\qquad
\left.\frac{\partial}{\partial N}\right|_{N=0}\hspace{-5mm}\Delta_T=-1/(9\pi)=-0.035367\cdots.
\label{derivatives2D}
\end{align}
It would be notable that this derivative for $\Delta_T$ for 2D is somehow almost unchanged in magnitude for 3D as we will see below.
The leading term in the $\epsilon$-expansion may be compared with \eqref{derivatives2D} as
\begin{align}
\left.\frac{\partial(\Delta_S-\Delta_T)}{\partial N}\right|_{N=0}=
\begin{cases}
1/\pi &       D=2,\\   
\epsilon/8  + \mathcal{O} (\epsilon^2)&  D=4-\epsilon,
\end{cases}
\end{align}
where each contribution of the derivative for $D=4-\epsilon$ is $\partial( \Delta_S , \Delta_T)/\partial N=(3\epsilon/32, -\epsilon/32)$.
In Appendix, we compute a pseudo $\epsilon$ series using the input of the 6-loops $D=3$ RG calculations \cite{Antonenko} and 
present reasonable estimates by a simple Pad\'e analysis 
together with the best-known results of the $\epsilon$-expansion up to $\epsilon^5$.  As a simple estimate, we take the average of the six and five-loops and the maximum deviation as an error. This gives,
\begin{align}
\left.\frac{\partial}{\partial N}\right|_{N=0}\hspace{-5mm}\Delta_S=0.1238(28),\qquad\qquad
\left.\frac{\partial}{\partial N}\right|_{N=0}\hspace{-5mm}\Delta_T=-0.036(7).
\label{derivativevalues0}
\end{align}
The same analysis for the derivatives at $N=1$ (the Ising point) yields
\begin{align}
\left.\frac{\partial}{\partial N}\right|_{N=1}\hspace{-5mm}\Delta_S=0.1017(35),\qquad\qquad
\left.\frac{\partial}{\partial N}\right|_{N=1}\hspace{-5mm}\Delta_T=-0.032(5),
\label{derivativevalues1}
\end{align}
The derivatives for $\Delta_T$, which are useful in this paper, increases only slightly ($\sim 10\%$) in the interval $N \in [0,1]$. 
Nevertheless, in order to get better estimates, one may found it more useful to keep both \eqref{derivativevalues0} and \eqref{derivativevalues1} than to choose one of these two. 
More concretely, the variation $\Delta_T(N_2)-\Delta_T(N_1)$ with $0\leqslant N_1<N_2\leqslant 1$ can be better approximated by 
$(N_2-N_1)$ times the derivative at the midpoint $(N_2+N_1)/2$ obtained as a linear interpolation between \eqref{derivativevalues0} and \eqref{derivativevalues1}.
For instance, a roughest estimate for the variation between $N=1$ and $N=0$ may be obtained as
$\Delta_T(1)-\Delta_T(0)=(+1)\times(-0.032(5)-0.036(7))/2=-0.034(4)$, where the errors in \eqref{derivativevalues0} and \eqref{derivativevalues1} are assumed to be independent. Although we do not use the last example, which would maximize the uncertainty, one may check
\footnote{This paper focuses on the approach to $N=0$ through $\Delta_T$. However, 
some crudest benchmark for the singlet dimension is possible using $\Delta_S(1)=1.41264(6)$ obtained for the Ising model \cite{Simmons-Duffin} as follows: 
$\Delta_S(0)=1.41264(6)+(-1)\times(0.1238(28)+0.1017(35))/2=1.300(22)$, which is consistent with $\Delta_T(0)$ in Section \ref{section:N=0} as expected.
}
this estimate may reasonably connect the results obtained independently by conformal bootstrap
in Section \ref{section:N=0} ($N=0$) and in Section \ref{section:N=1} ($N=1$).


\section{Operator product expansion of the fundamental fields in the $O(N)$ CFT}


\subsection{Crossing symmetry sum rule}\label{section:sumrule}

The crossing symmetry sum rule used in this paper is the most basic one 
(in the sense it does not involve the mixed correlators \cite{Kos15})
in the conformal bootstrap for the 
CFT with a global $O(N)$ symmetry \cite{Kos14,Rattazzi,Vichi,Poland} as described briefly below. 
The fundamental field in this theory is a scalar operator $\phi_a$, which transforms as an $O(N)$-vector, with dimension $\Delta_{\phi}$.
Crucially, the OPE of $\phi_a$ with itself may be decomposed into three sectors:
\begin{align}
\label{OPE}
\phi_a(x) \times \phi_b(0) \sim \sum_{(\Delta, \ell) \in S} \lambda^S_{ \Delta, \ell} \cO^{S, \Delta, \ell} \delta_{ab}+ 
\sum_{(\Delta, \ell) \in T} \lambda^T_{ \Delta, \ell}\cO_{(ab)}^{T, \Delta, \ell} + 
\sum_{(\Delta, \ell) \in A} \lambda^A_{\Delta, \ell}\cO_{[ab]}^{A, \Delta, \ell},
\end{align}
where $S$, $T$, and $A$ denote 
the $O(N)$ singlets sector of even spin, 
the $O(N)$ symmetric tensor sector of even spin, and
the $O(N)$ anti-symmetric tensor sector of odd spin, respectively.
Note that there is an infinite tower of the scaling dimensions $\{\Delta_{\ell}^{(0)}, \Delta_{\ell}^{(1)}, \cdots\}$ for the states with fixed $\ell$ in each sector.  
The dependence on $x$ is omitted on the right hand side. 
The sets of the OPE coefficients $\lambda_{\Delta, \ell}^X = \lambda_{\phi\phi}^{\cO^{X, \Delta, \ell}}$ 
(with $X=S,\,T,\,A$ and the tensor labels are omitted) 
encode important dynamical information in the $O(N)$ CFT
and satisfy highly nontrivial constraints due to the associativity of the operator algebra. 
These constraints can be expressed as a sum rule that follows from the equivalence (the crossing symmetry) of the two different expansions of a single four point function
$\left\langle\phi_a(x_1)\phi_b(x_2)\phi_c(x_3)\phi_d(x_4)
\right\rangle$
from the two distinct degeneration limits ($x_1\to x_2$ and $x_1\to x_4$), 
where the contribution from the identity operator, which belongs to the singlet ($S$) sector, becomes dominant.

Let us write the contribution from each sector in the OPE \eqref{OPE} in the channel $x_1\to x_2$ as follows:
\begin{align}\label{shorthand}
\mathcal{S}'\equiv\sum_{(\Delta,\ell)\in S'} \lambda_{\Delta,\ell}^2 G_{\Delta, \ell}(u,v),~~ 
\mathcal{T}\equiv\sum_{(\Delta,\ell)\in T} \lambda_{\Delta,\ell}^2 G_{\Delta, \ell}(u,v),~~
\mathcal{A}\equiv\sum_{(\Delta,\ell)\in A} \lambda_{\Delta,\ell}^2 G_{\Delta, \ell}(u,v).
\end{align}
Here in the first sum, the set $S'$ includes all the operator in $S$-sector except the the identity operator $(\Delta, \ell)=(0,0)$,
for which the contribution for the four-point function is simply $(x_{12}^2x_{34}^2)^{-\Delta_\phi}\delta_{ab}\delta_{cd}$,
which is usualy the dominant contribution in the limit $x_1\to x_2$.  
For concreteness, we note that the conformal partial wave (global conformal blocks) $G_{\Delta, \ell}(u,v)$ in $D$-dimensions is a function of the  
two cross-ratios given as
\begin{align}
u = z \bar{z} = \frac{x_{12}^2 x_{34}^2}{x_{13}^2 x_{24}^2},  \qquad 
v = (1-z)(1-\bar{z}) = \frac{x_{14}^2 x_{23}^2}{x_{13}^2 x_{24}^2},  \qquad x_{ij} = x_i - x_j,
\end{align}
and has the following form \cite{Hogervorst} in terms of the radial coordinates $r e^{i\theta}=z/(1+\sqrt{1-z})^2$ 
\begin{align}\label{globalblock}
G_{\Delta,\ell}(r,\theta)=\sum_{n=0}^{\infty}\sum_{j} B_{n,j}^{(\ell)} r^{\Delta +n} C^{\nu}_j (\cos\theta),
\end{align}
where the coefficients $B_{n,j}^{(\ell)}$ with $j=\ell+n, \ell+n-2, \cdots, \max(\ell-n, \frac{1+(-1)^{\ell+n+1}}{2})$ can be iteratively fixed by the Casimir differential equation for the $D$-dimensional conformal group, 
and $C^{\nu}_j$ with $\nu=(D-2)/2$ is the Gegenbauer polynomial.

The crossing symmetry can be nicely seen using the OPE \eqref{OPE} and the notation \eqref{shorthand} as follows:
\begin{align}
\left\langle
\begin{array}{cc}
\phi_a(x_1) & \phi_d(x_4) \\
\phi_b(x_2) & \phi_c(x_3) \\
\end{array}
\right\rangle
=&\frac{1}{(x_{12}^2x_{34}^2)^{\Delta_\phi}} 
\left\{ )(~\left( 1+\mathcal{S}'\right) + \Bigl[  \asymp +  \slash\hspace{-2.5mm}\setminus - \frac{2}{N} )( \Bigr] \mathcal{T} + \Bigl[ \asymp - \slash\hspace{-1.8mm}\setminus \Bigr] \mathcal{A}    
\right\}\\
=&\frac{1}{(x_{12}^2x_{34}^2)^{\Delta_\phi}} \left(\frac{u}{v} \right)^{\Delta_\phi}
\left\{ \asymp~\left( 1+\tilde{\mathcal{S}}'\right) + \Bigl[ )( + \slash\hspace{-2.5mm}\setminus -\frac{2}{N} \asymp \Bigr] \tilde{\mathcal{T}} + \Bigl[ )( - \slash\hspace{-1.8mm}\setminus  \Bigr] \tilde{\mathcal{A}}    
\right\}
\end{align}
where all the possible three tensor structures after the contractions are represented as
$)(=\delta_{ab}\delta_{cd}$, 
$\slash\hspace{-1.7mm}\setminus$$=\delta_{ac}\delta_{bd}$, 
$\asymp=\delta_{ad}\delta_{bc}$, and 
the tilde notation is used for representing the quantities with $u$ and $v$ interchanged, thus
$\mathcal{S}'\equiv\mathcal{S}'(u,v)$, $\tilde{\mathcal{S}}'\equiv\mathcal{S}'(v,u)$, etc..
Using the notation $\mathcal{X}_{\pm}\equiv v^{-\Delta_\phi}\mathcal{X}  \pm u^{-\Delta_\phi}\tilde{\mathcal{X}}$ 
($\mathcal{X}=1, \mathcal{S}'$, $\mathcal{T}$, $\mathcal{A}$), 
the $O(N)$ sum rule follows by comparing the terms for each tensor structure $\asymp$, $)($, and  $\slash\hspace{-1.8mm}\setminus$:
\begin{align}
\mathcal{S}'_{-}+\left(1-\frac{2}{N}\right)\mathcal{T}_{-} + \mathcal{A}_{-}&=-1_{-}, \label{onsum1}\\
\mathcal{S}'_{+}-\left(1+\frac{2}{N}\right)\mathcal{T}_{+} - \mathcal{A}_{+}&=-1_{+}, \label{onsum2}\\
\mathcal{T}_ {-}-\mathcal{A}_{-}&=0. \label{onsum3}
\end{align}
For each $N$, the solution manifold for the crossing symmetry \eqref{onsum1}-\eqref{onsum3} consisting of the points represented by the effective CFT data 
(the possible set of scaling dimensions and spins $(\Delta, \ell)$ in $X$-sector 
with associated OPE coefficients $\lambda_{\Delta, \ell}^X$) may be inifinite dimensional.
An important one-parameter-family solution, which is conveniently parametrized by $\Delta_\phi$, 
can be singled out along the boundary of the unitarity (dictated by the lower bounds \eqref{unitaritybound} and the positivity \eqref{positivity}),
whose projection onto $\Delta_\phi$-$\Delta_T$ plane is shown for each $N$ in FIG. \ref{fig:kink}.
As is well-known, the search for this unitarity saturating solution can be formulated as a linear optimization problem (see Section \ref{section:bootstrap} for more details)
and can be solved with the aid of knowledge on the global conformal blocks  $G_{\Delta,\ell}(u,v)$ such as that in \eqref{globalblock}.


In order to clear up a common source of confusion, 
it is worth to make a careful distinction between the spectrum of the $O(1)$ model ($N=1$) and 
that of the Ising model in our formulation.
The Ising ($\mathbb{Z}_2$) sum rule, which is used in \cite{El-Showk} for instance,
follows from the crossing symmetry of the four-point function
for a single scalar $\left\langle\phi(x_1)\phi(x_2)\phi(x_3)\phi(x_4)\right\rangle$.
Since only the singlet fields appear in the Ising OPE $\phi\times\phi$, the whole contribution in $x_1\to x_2$ except that from the identity operator may be denoted as $\mathcal{S}'$. 
Then the crossing symmetry leads to 
$1+\mathcal{S}'=(u/v)^{\Delta_\phi} (1+\tilde{\mathcal{S}}')$, which further simplifies to,
\begin{equation}
\mathcal{S}'_{-}=-1_{-}.
\label{isingsum}
\end{equation}
Now the logic is as follows. A proper subset  \eqref{onsum1} and \eqref{onsum3} of the $O(N)$ sum rule for $N=1$ implies the Ising sum rule \eqref{isingsum}, which, along with the requirements of saturating unitarity 
(\eqref{unitaritybound} and \eqref{positivity}), is sufficient for a given $\Delta_\phi$ to single out a unique solution for $\mathcal{S}'$. 
Thus, in particular, the unitarity-saturating solution for the Ising sum rule may be embedded into the solution for the $O(1)$ sum rule as its $S$-sector. 
In this case, one may still generalize this Ising spectrum to a solution for the $O(1)$ sum rule,
which may also admit non-empty $T\oplus A$ sectors, in addition to $S$-sector, determined in turn by solving
$-3\mathcal{T}_{+} - \mathcal{A}_{+}=-1_{+}-\mathcal{S}'_{+}$ and
$\mathcal{T}_ {-}-\mathcal{A}_{-}=0$, where the Ising contribution $\mathcal{S}'_{+}$ 
may be regarded as a seed generating these sectors.
In particular, $T$-sector in this solution contains the rank-2 symmetric tensor operator $\varphi_{ab}$ in \eqref{phi_ab} with a non-vanishing squared OPE coefficient also for $N=1$,  
which actually determines the fractal dimensions for the Ising high temperature graphs as shown in Section \ref{section:N=1}.

\subsection{The central charges $C_T$ and $C_J$}\label{section:cTcJ}
If we omit RG irrelevant operators and keep only most important ones, the OPE \eqref{OPE} becomes
\begin{align}
\label{operatorcontent}
\phi_{a}\times \phi_{b}  = 
\left( {\bf 1} + \lambda^S_{\Delta_S, 0}~\varepsilon
+ \lambda^S_{D, 2}~T^{\mu\nu}\right)\delta_{ab} 
+ \lambda^T_{\Delta_T, 0}~\varphi_{(ab)}+ 
   \lambda^A_{D-1, 1}~ J^{\mu}_{[ab]} +\cdots, 
\end{align}
where $T^{\mu\nu}$ and $J^{\mu}_{ab}$ are the stress-energy tensor and the conserved vector current, respectively.
As the symbol in \eqref{operatorcontent} signifies, $T^{\mu\nu}$ is a spin-2 $O(N)$ singlet with dimension $D$ and 
$J^{\mu}_{ab}$ is a spin-1 anti-symmetric vector ($J^{\mu}_{ab}=-J^{\mu}_{ba}$) with dimension $D-1$, 
which transform as an $O(N)$-adjoint representation.
The conformally invariant two-point functions of
$T^{\mu\nu}(x)$ and $J^{\mu}_{ab}(x)$ are \cite{Petkou96},
\begin{align}\label{centraldefinition}
\left\langle T^{\mu\nu}(x_{1})T^{\rho\sigma}(x_{2})\right\rangle  = 
\frac{C_{T}}{S_D^2}~\frac{I^{\mu\nu ,\rho\sigma}(x_{12})}{x_{12}^{2D}}, \qquad
\left\langle J^{\mu}_{ab}(x_{1})J^{\nu}_{cd}(x_{2})\right\rangle  = 
\frac{C_{J}}{S_D^2}\frac{I^{\mu\nu}(x_{12})}{x_{12}^{2(D-1)}}~(\delta_{ac}\delta_{bd}-\delta_{ad}\delta_{bc}),
\end{align}
with normalization given by the surface of unit $(D-1)$-sphere $S_D=2\pi^{D/2}/\Gamma(D/2)$ and with
\begin{align}
I^{\mu\nu ,\rho\sigma}(x)  =  
\frac{1}{2}\Bigl(I^{\mu\rho}(x)I^{\nu\sigma}(x)+I^{\mu\sigma}(x)I^{\nu\rho}(x)\Bigl)
\,-\,\frac{1}{D}\delta^{\mu\nu}\delta^{\rho\sigma}, \qquad
I^{\mu\nu}(x)  = 
\delta^{\mu\nu}-2\frac{x^{\mu}x^{\nu}}{x^{2}}.
\end{align}
The Ward identities for the stress-energy tensor $T$ and the conserved current $J$ leads to
\begin{align}\label{wardidentities}
NC_{T}/C_{T; \mathrm{free}}=\Delta_\phi^2/\left(\lambda^S_{D, 2}\right)^2, \qquad C_{J}/C_{J; \mathrm{free}}=1/\left(\lambda^A_{D-1, 1}\right)^2,
\end{align}
where
$C_{T, \mathrm{free}}=ND/(D-1)$ and 
$C_{J, \mathrm{free}}=2/(D-2)$ are free field values.
For the $O(N)$ CFT, some useful results are known in the IR fixed point. These include the $\epsilon=4-D$ expansion
\begin{align}
\label{centralepsilon}
C_{T}/C_{T; \mathrm{free}}=1 - \frac{5}{12}\frac{N+2}{(N+8)^2}\epsilon^2 + \mathcal{O}(\epsilon^3), \qquad 
C_{J}/C_{J; \mathrm{free}} =1 - \frac{3}{4}\frac{N+2}{(N+8)^2}\epsilon^2 + \mathcal{O}(\epsilon^3),
\end{align}
and the $1/N$-expansion in $D=3$
\footnote{For $1/N$ coefficient, there is a mismatch by a factor 2 between (4.25) and (6.8) of \cite{Petkou96}. 
Our results on the slope \eqref{central_delta} 
in FIG. \ref{fig:cJ} as well as \cite{Cha,Huh} supports the value $-64/9\pi^2$ in \eqref{central1N} reproduced from (4.25). 
Also by using a Pad\'{e} analysis on \eqref{phi1N},  the universal curves 
$\Delta_\phi$-$\Delta_S$ and $\Delta_\phi$-$\Delta_T$ in FIG. \ref{fig:universalcurves} can be drawn,
which will be discussed elsewhere.} 
\begin{align}
\label{central1N}
C_{T} /C_{T; \mathrm{free\,}O(N)}=1 - \frac{40}{9\pi^2}\frac{1}{N} + \mathcal{O}\left(\frac{1}{N^2}\right), \qquad 
C_{J} /C_{J; \mathrm{free}} =1 - \frac{64}{9\pi^2}\frac{1}{N} + \mathcal{O}\left(\frac{1}{N^2}\right).
\end{align}
Regarding \eqref{centralepsilon} and \eqref{central1N}, it is an open direction to study \cite{Petkou96,Vilasis} under which conditions 
these $C_T$ and $C_J$ in 3D are monotonically decreasing along the RG as the central charge does so in the 2D unitary system
\cite{Zamolodchikov}. 
Together with the leading correction in $1/N$ in the expansion \cite{Moshe},
\begin{align}
\label{phi1N}
\Delta_\phi=\frac{1}{2} + \frac{4}{3\pi^2 N} - \frac{256}{27\pi^2 N^2}
+\frac{32 \left(3 \pi ^2 (-3402 \zeta (3)-61+108 \log (2))-3188\right)}{243 \pi ^6 N^3}
+\mathcal{O}(\frac{1}{N^4})
\end{align}
the large $N$ asymptotics of the central charges $C_{T}$ and $C_{J}$
may be written as a function of $\Delta_\phi$ as
\begin{align}
\label{central_delta}
C_{T}/C_{T; \mathrm{free}} = 1 - \frac{10}{3}\Delta_\phi + \mathcal{O}\left( \Delta_\phi^2 \right), \qquad 
C_{J}  = 2 - \frac{32}{3}\Delta_\phi + \mathcal{O}\left( \Delta_\phi^2 \right),
\end{align}
where the slope $\partial C_T/\partial\Delta_\phi$ has been already considered in \cite{Kos14}.
On the other hand,
we observe that the other one $\partial C_J/\partial\Delta_\phi$ shows interesting behavior (FIG. \ref{fig:cJ}) for the case $N\sim 0\ll 1$ of our main interest
\footnote{We thank Tomoki Ohtsuki for pointing out that 
similar kinks in $C_J$ can be observed via the direct $C_J$ minimization in $D=3$ \cite{Nakayama} for $N\geqslant 2$, and Yu Nakayama for further discussions. 
For $C_T$ in 3D, it is known that the direct $C_T$ minimization reproduces $C_T$ along the unitarity bound (via $\Delta_{S}$-maximization) for $N=1$, 
but not for generic $N>2$ \cite{Kos14}. It is possible to check similar characteristics are shared by $C_J$ in 3D. It would be also interesting to study the implication of these phenomena on the solution space of the crossing symmetry.}.
The formation of an effective minimum near $N\sim 0$ is discussed in Section \ref{section:slopechange}
and used to estimate the fractal dimension $d_F$ of polymer chains in Section \ref{section:N=0}.

\subsection{Conformal bootstrap, the unitarity bound, and the search space}\label{section:bootstrap}
A particularly important one-parameter-family solution of the crossing symmetry \eqref{onsum1}-\eqref{onsum3} lies along the boundary of unitarity, 
which may 
connect the whole spectrum of the free theory and the $O(N)$ CFT.
This solution can be singled out by a linear optimization as mentioned in the end of Section \ref{section:sumrule}.
The unitarity consists of the following two conditions.
First, the scaling dimensions in a $D$-dimensional theory must satisfy the lower bounds, which correspond to the 
requiment that the anomalous dimensions be positive) \cite{Ferrara,Mack,Metsaev,Minwalla}, 
\begin{equation}
\label{unitaritybound}
\Delta \geqslant 
\begin{cases}
\ell+D-2 & \text{for}\qquad \ell > 0, \\
\frac{D-2}{2}  & \text{for}\qquad \ell = 0,
\end{cases}
\end{equation}
where the inequalities are saturated by conserved currents such as $T^{\mu\nu}$ and $J^{\mu}_{ab}$ ($\ell>0$)
and by free scalars such as a fundamental field $\phi_a$ at the free theory ($\ell=0$).
Second, the squared OPE coefficients must be positive:
\begin{align}
\label{positivity}
\left(\lambda^{\cO^{X, \Delta, \ell}}_{\phi\phi}\right)^2 > 0 \quad \text{for all the operators } \cO^{X, \Delta, \ell} \text{ in \eqref{OPE}}.
\end{align}
These two requirements of the unitarity enable one to solve the crossing symmetry \eqref{onsum1}-\eqref{onsum3} along 
the unitarity bound via the simplex algorithm \cite{El-Showk,El-Showk14,Paulos} or the semi-definite program \cite{Simmons-Duffin}.
 
We use the standard simplex algorithm (Sec. 6 of \cite{El-Showk14}) with our particular implementation based on the code \cite{Paulos}.
As usual, the simplex algorithm is used in order to try to determine if there exists a solution of 
the crossing symmetry \eqref{onsum1}-\eqref{onsum3} that satisfies the lower bounds \eqref{unitaritybound} and 
the positivity \eqref{positivity}
for a given $\Delta_\phi$ and in the region $\Delta_X> \Delta_{X_0}$ for the dimension $\Delta_{X_0}$ of some low-lying operator;
here we use it for the symmetric tensor $\varphi_{ab}$ ($\Delta_X=\Delta_T$) mainly for the reason given in Section \ref{section:kink}-{\it b}.
If solutions do not exist (do exist), the next search region $\Delta_T>\Delta_{T_1}$ can be chosen narrower
such that  $\Delta_{T_1}<\Delta_{T_0}$ ($\Delta_{T_1}>\Delta_{T_0}$).
Then one may take a bisection procedure from some initial finite interval of $\Delta_T$ and narrow the search region by each trial. 
If the initial interval is taken wide enough, the iteration eventually reaches the upper-bound $\Delta_{T_\infty}$ for $\Delta_{T}$, 
for which the solution in $\Delta_{T} \geqslant \Delta_{T_\infty}$ is expected to be unique (in particular, $\Delta_{T} = \Delta_{T_\infty}$).
The value of $\Delta_{T_\infty}$ is measured numerically by setting the bisection accuracy goal $\delta(\Delta_T)$, 
which we typically take $\delta(\Delta_T)$ less than $10^{-4}$. 

In practice, the crossing symmetry constraints are extracted by a truncated Taylor-expansion around the symmetric point $u=v=1/4$ 
of the sum rule \eqref{onsum1}-\eqref{onsum3} also with a truncated number $\ell_{\text{max}}$ of the spin sectors.
The simplex algorithm (at $j$-th step of the bisection) searches the spectrum region bounded from below 
by \eqref{unitaritybound} (with $\Delta_T\geqslant \Delta_{T_{j-1}}$) and from above by an appropriate upper bound $\Delta_{\text{max}}$,
which should be taken large enough.
The derivatives of \eqref{onsum1}-\eqref{onsum3} are computed with respect to the coordinate $(a,b)$ defined from 
$(z, \bar{z})=(a+\sqrt{b},a-\sqrt{b})/2$, which is related to the cross-ratios as $(u,v)=(z\bar{z}, (1-z)(1-\bar{z}))$.
Following the convention in \cite{Paulos}, they are reduced to the following set of the derivatives of the global conformal block $G_{\Delta, \ell}(a,b)$ in \eqref{globalblock},
which one may select as  
\begin{equation}
\label{derivativechoice}
\left\{ \partial_a^m\partial_b^n G_{\Delta, \ell}(a=1, b=0)
~\Bigr|  \quad m=0,\, \cdots,\, 2(n_{\text{max}} -n) +m_{\text{max}};\quad n=0,\,\cdots ,\,  n_{\text{max}}  \right\}
\end{equation}
with some $(m_{\text{max}}, n_{\text{max}})$, which consists of $\mathcal{K}=(m_{\text{max}}+n_{\text{max}}+1)(n_{\text{max}}+1)$ derivatives.
In general, the unitarity bound becomes more strict for larger number $\mathcal{K}$ of the derivatives, 
although an exact form of convergence to the optimal bound is not well-understood so far.
Also for a given $\mathcal{K}_0$, the bound usually depends only scarcely to a paticular choice of $(m_{\text{max}}, n_{\text{max}})$ 
with $\mathcal{K}\sim \mathcal{K}_0$.
The number of spins $\ell_{\text{max}}$ should be taken large enough with respect to the choice $(m_{\text{max}}, n_{\text{max}})$
so that resulting bound does not depend on $\ell_{\text{max}}$. 
Our default choice used for measuring the critical exponents
is $(m_{\text{max}}, n_{\text{max}})=(8,8)$ with $\mathcal{K}=153$, $\Delta_{\text{max}}=70$, and $\ell_{\text{max}}\sim 50$.
 
\section{Singular shapes of the unitarity bounds}\label{section:singularshape}
\subsection{Smoothing of the kink in the unitarity bound for the tensor dimension $\Delta_T$}\label{section:kink}
A crucial observation that gave an initial momentum to the recent revival of the bootstrap studies
was perhaps the emergence of the singular shape (a kink) at $\Delta_\phi=\Delta_{\phi;\; \text{Ising}}$ in the unitarity bound curve of 
the scaling dimension $\Delta_{\phi^2}=\Delta_{\phi^2}(\Delta_{\phi})$ 
in the case of $\mathbb{Z}_2$ symmetry studied for the Ising model
\footnote{In view of the picture that there are infinitely many $\mathbb{Z}_2$ symmetric primary operators above $\varepsilon=:\phi^2:$ whose levels 
are separated by non-trivial intervals and may be repulsive to each other, the observed straightness of the lowest level $\Delta_{\phi^2}$ on 
the right side of the kink ($\Delta_\phi>\Delta_{\phi;\; \text{Ising}}$) is also remarkable.}
.
Actually, the appearance of singular shapes in the conformal bootstrap seems more ubiquitous as we will see also in this work.

Below we use the sum rule \eqref{onsum1}-\eqref{onsum3} for non-integer $N$.
It is important to realize that the unitarity is violated in the free $O(N)$ models for any non-integer $N$ \cite{Maldacena}, and 
it is also most likely to be so in the IR fixed point of the 3D $O(N)$ model for non-integer $N$.
In particular, one should not expect the IR fixed point lies exactly at the unitarity saturating solutions.
Nevertheless, the results seem to point to a weak effect of violation  
that the IR fixed point may still lie close to the unitarity-saturating solution
as long as $N$ is positive and is not too close to $N=0$. 
For $N<0$, a severer effect of the violation prevents one from obtaining meaningful bounds as discussed in the end of Section 
\ref{section:convergence}.
More discussions on the effect of the unitarity violation for generic non-integer $N>0$ is postponed to Section \ref{section:conclusion}.

\begin{figure}
\includegraphics[width=12cm]{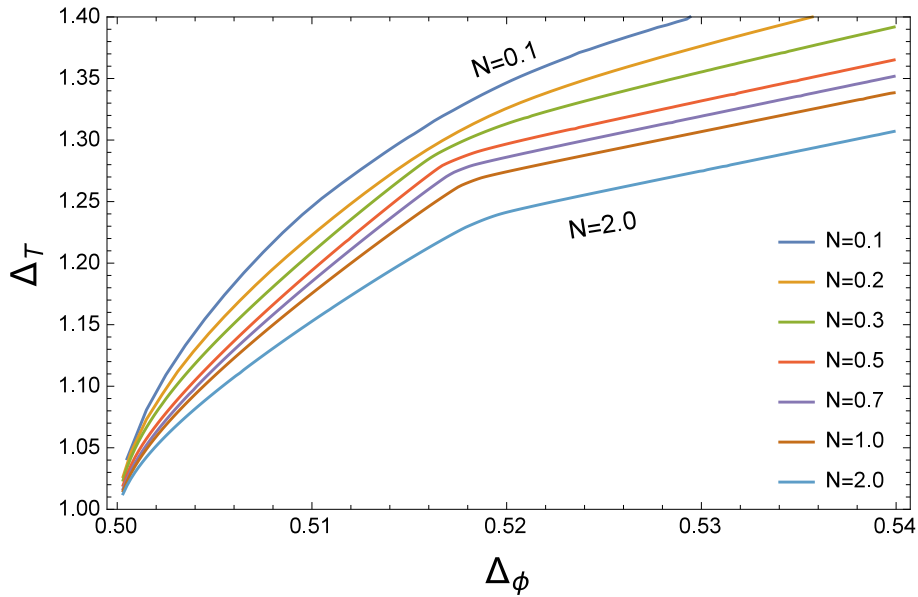} 
	\caption{\footnotesize
      \label{fig:kink}
      The unitarity upper bound for the dimension $\Delta_T$ of the tensor $\varphi_{ab}$
obtained with the derivative cut off set at 
$(n_{\text{max}}, m_{\text{max}})=(8,8)$ for $N=0.1$ and 
$(n_{\text{max}}, m_{\text{max}})=(7,2)$ otherwise.
We have also verified that a higher cut off makes the kinks sharper for generic $N$, 
but not so effectively for $N\leqslant 0.2$.        
} 
\end{figure}

We here show the unitarity upper-bound obtained by the bisection for $\Delta_T$ (Section \ref{section:bootstrap})  
in the $O(N)$ model with $0.1 \leqslant  N \leqslant2$ in FIG. \ref{fig:kink}. 
This result may be regarded as an extention to the region $N< 2$ of the bounds previously obtained for $\Delta_{T}$ for $N\geqslant 2$ (and $\Delta_{S}$ for $N\geqslant 1$) \cite{Kos14}, where observed single kink has been used to estimate the scaling dimensions $(\Delta_\phi, \Delta_Y)$ in the $O(N)$ CFT for each $N$ and for each sector $Y=S, T$.
As one can see, the curves in FIG. \ref{fig:kink} for $N\geqslant 0.5$ has a clearer kink and these can be used to estimate $(\Delta_\phi, \Delta_T)$ of the $O(N)$ CFT; 
note that once $\Delta_\phi$ is determined,  other low-lying dimensions are also available 
since the simplex algorithm has reached a unique solution  (see the end of Section \ref{section:sumrule}) 
for the crossing symmetry \eqref{onsum1}-\eqref{onsum3} along the unitarity bound for a given $\Delta_\phi$. 
 
Remarkably, however, the kink becomes less and less pronounced as $N$ tends to zero (the polymer limit)  
thus practically making the determination of $\Delta_\phi$ more difficult. 
Even in such circumstances, other singular shapes may remain in other universal quantities along the unitarity bound. 
This is indeed the case, and we observed a very clear change of the slope $\partial C_J/\partial\Delta_\phi$ in the current central charge $C_J$
as discussed in Section \ref{section:slopechange} and used this for the limit $N\to 0$ in Section \ref{section:N=0}.
Now two important remarks are in order. 

\subsubsection{Convergence to the optimal bound.}\label{section:convergence}
The first is on the convergence of the bound to the optimal shape
with respect to the truncation \eqref{derivativechoice} of the derivative orders $(m, n)$ 
on the conformal block $G_{\Delta, \ell}(a,b)$, which is currently unavoidable in numerics.
Although the upper-bound obtained in a finite truncation is rigorous, 
a larger number $\mathcal{K}$ of derivatives leads to more restrictive bound (i.e. a lower upper-bound) 
and actually makes the kink more sharp, where the convergence to the optimal bound tends to be faster than that in the rest.  
In addition to this, the convergence becomes much slower in the polymer limit $N\to 0$.
Thus a bruteforce approach to the limit is taking $\mathcal{K}$ large enough with respect to a given small $N$. 
In practice, the unitarity bound for $\Delta_T$ looks smooth for $N\lesssim 0.2$, 
which makes the detection of the kink (a discontinuity in the slope) becomes very hard 
within a reasonablely large number of derivatives ($\mathcal{K}=153$).
Here it would be also worth noting the present guess on the optimal shape  ($\mathcal{K}=\infty$) for the limit $N\to 0$.
On the right of the kink, the convergence of the slope $\partial C_T/\partial\Delta_\phi$ to a almost constant, as in the cuves for $N\geqslant 0.5$,  seems to be plausible. 
On the left, finite-$\mathcal{K}$ curves are convex upward as in FIG. \ref{fig:kink}.
It seems, however, there are no strong indications that excludes the possiblity that the optimal shape on the left is also almost straight in 3D, 
while there is a 2D example where the bound is likely to be convex downward \cite{Rychkov}.
In general, it would be interesting to consider if there is a principle that forbids the optimal unitiarity bound for these scaling dimensions be convex upward with some reasonable assumptions.
We will give a qualitative argument in Section \ref{section:slopechange} on the enhanced slope $\partial \Delta_T/\partial{\Delta_\phi}$ on the left of the kink 
in view of the level dynamics.

The slowdown of the convergence is probably related to the degeneracy of two levels $\Delta_T=\Delta_S$ 
and {\it  the severe unitarity wall} at $N=0$, where some squared OPE coefficients 
\footnote{Another important example of a singlular OPE coefficient in $N\to 0$ may be $\lambda^\varepsilon_{\varepsilon\varepsilon}$ 
for three energy operators, which would play a role in the mixed correlator bootstrap \cite{Kos15}. 
The physical origin of the divergence of $\lambda^\varepsilon_{\varepsilon\varepsilon}$ can be traced back to the strong repulsion between 
the loop segments $\varepsilon$ in the $O(N\to 0)$ loop model \cite{Shimada}.}
including $(\lambda^S_{D, 2})^2$ for the stress-energy tensor has a pole, which can be seen by the relation \eqref{wardidentities} from the Ward identity and 
by our observation that the ratio $C_{T}/C_{T; \mathrm{free}}$ remains finite $\sim 0.955$.
The detailed analysis on this special limit $N\to 0$ and 
quantitative knowledge on the order of convergence in the conformal bootstrap 
in general might be useful and would deserve further investigation.

\subsubsection{Gap assumptions.}\label{gapassumption}
 The second remark is especially relevant if one tries to find the upper bound for $\Delta_{S}$ by the $\Delta_{S}$-bisection in the case $N< 1$.
In \cite{Kos14}, it was found useful to complement the unitarity lower bounds \eqref{unitaritybound} by an extra gap assumption 
that scalar fields ($\ell=0$) in the right hand side of the OPE \eqref{OPE} 
be bounded below by the RG canonical dimension: 
\begin{equation}
\label{canonicaldimension}
\Delta \geqslant D-2 \qquad \text{for } \ell = 0 \text{ operators in the R.H.S. of the OPE \eqref{OPE}}. 
\end{equation}
This has previously been used  in order just to improve numerical stability \cite{Kos14}; in particular,
the condition $\Delta_T\geqslant 1$ in $D=3$ was not supposed to change the resulting solution of \eqref{onsum1}-\eqref{onsum3} through the $\Delta_{S}$-bisection.
Indeed, it can be checked that this gap assumption makes no distinction in the resulting spectrum for $N\geqslant 2$.

In the bootstrap for $N<1$, however, we found that the $\Delta_{S}$-bisection using the pure unitarity condition 
\eqref{unitaritybound} and \eqref{positivity}  may yield a solution  
that violates this additional gap assumption \eqref{canonicaldimension}, namely, a solution with $1/2<\Delta_T<1<\Delta_S$,
which satisfies \eqref{unitaritybound}, but can not correspond to the $O(N)$ CFT from the RG point of view.
This phenomenon may remind us of a level repulsion between $\Delta_T$ and $\Delta_S$ in the solution space that becomes stronger as $N\to 0$.
Accordingly, if one sticks to keep the extra condition $\Delta_T\geqslant 1$,
the $\Delta_{S}$-bisection yields another unphysical solution 
\footnote{Around $N=1$ the saturation of \eqref{canonicaldimension} may not be so serious as the kink in $\Delta_{S}$ appears around the expected Ising position, 
which should be consistent with our observation that  the OPE coefficient $\lambda^T_{1,0}$ of the (unphysical) level $\Delta_T=1$ is negligible compared to those for other operators.
However, below $N=1$ this makes much difference: for instance, a kink in $\Delta_S$ emerges  even in $N=0.1$, 
which was smoothed in the solution with the pure unitarity conditions. 
Again, it is obvious that this solution with $\Delta_T=1$ can not represent a physical spectrum.} 
which contains $\Delta_T=1<\Delta_S$ and thus saturates the extra gap assumption set by hand.
In contrast to the $\Delta_{S}$-bisection, our bisection for $\Delta_{T}$, which should be 
the lowest dimension scalar contained in the product $\phi_a\times \phi_b$ in the $O(N)$ CFT with $N>0$, is free from such problems.
In particular, the extra assumption $\Delta_S\geqslant 1$ does not change the resulting solution.
As we have just seen, besides the direct role of determining the fractal dimension \eqref{d_F}, 
the property that $\Delta_T$ has the lowest dimensions (like a ground state in quantum mechanics) 
in the right hand side of the OPE \eqref{OPE} adds the study of $\Delta_T$ a special importance.

\subsection{The slope change in $C_J$ and the level degeneracy in $A$-sector}\label{section:slopechange}

\begin{figure}
\includegraphics[width=16cm]{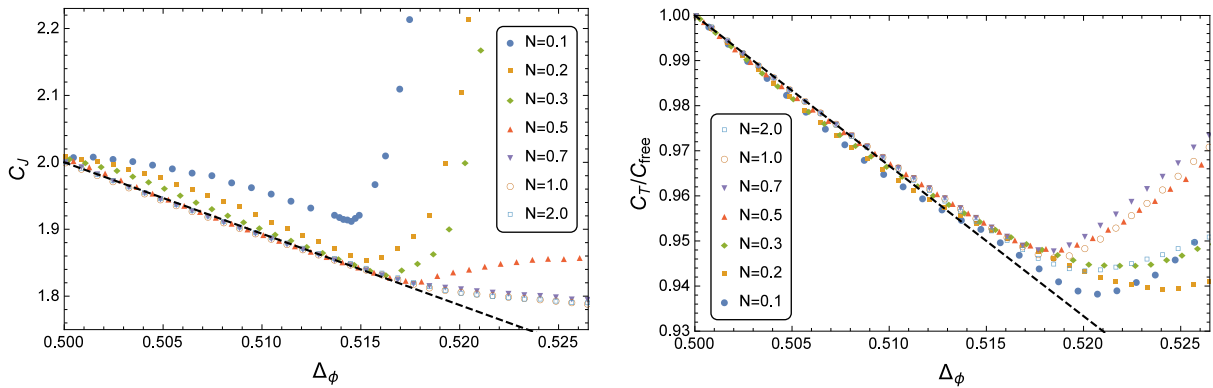} 
	\caption{\footnotesize
      \label{fig:cJ}
      The current central charge $C_J$ (left) and the central charge $C_T$ (right) as functions of $\Delta_\phi$.
The curves are obtained with the derivative order cut off at $(n_{\text{max}}, m_{\text{max}})=(8,8)$ for $N=0.1$ and at $(n_{\text{max}}, m_{\text{max}})=(7,2)$ otherwise. 
The large $N$ asymptotics \eqref{central_delta} with the slopes $-32/3$ and $-10/3$ are also shown (dashed).
    } 
\end{figure}

The singular shape is not restricted to the unitarity bound for scaling dimensions;
it may also appear in the unitariy bound for the OPE coefficients of the conserved currents such as $T^{\mu\nu}$ and $J^{\mu}_{ab}$, which are via the Ward identities reflected on sudden changes of the slopes in the central charge $C_T$ and the current central charge $C_J$ defined in \eqref{centraldefinition}.
We show $C_T/(NC_{\text{free}})$ and $C_J$ for $0.1\leqslant N \leqslant 2$ in FIG. \ref{fig:cJ},
most of which has just one point where the slope change occurs.
As the interaction in the $O(N)$ CFT becomes infinitesimal in the $N\to \infty$ limit, both changes in $(\Delta_\phi, C_T)$ and $(\Delta_\phi, C_J)$ 
from the free field values tends to zero. This corresponds to the asymptotics of these central charges given by \eqref{central_delta}; 
for a finite $N$, not too small, these slopes are actually shared as the initial slopes for small $\Delta_\phi -1/2$, where the effective interaction may be weak.     
In the case of our interest ($N\to 0$), a change of the slope $\partial C_J/\partial\Delta_\phi$ is enhanced and 
an effective minimum is formed for $N<0.5$. 
The effective minimum is used to estimate the dimension $\Delta_\phi$ of the fundamental field in $N\to 0$ in Section \ref{section:N=0}.
Before this application, we present some preliminary analysis on the mechanism
 ({\it a level dynamics} along the unitarity saturating solution that 
connects the free field theory and the $O(N)$ CFT, where the levels are essentially eigenvalues of the infinite dimensional matrix 
obtained by linearizing the RG flow around the fixed point in the theory space) 
behind the formation of these kinks.

As $\Delta_\phi =1/2$ corresponds to the free field theory, the effective anomalous dimension $\eta=2(\Delta_\phi-1/2)$ 
may be considered as an effective interaction parameter. 
If one traces the spectrum of scaling dimensions along the unitarity bound, one will meet a reorganization of the spectrum 
when $\Delta_\phi$ crosses the value at the kink ${\Delta_\phi}^*$, which is expected to be ${\Delta_\phi}$ of the $O(N)$ CFT 
as a similar phenomenon has been observed in the Ising model \cite{El-Showk14}. 
Along the unitarity-saturating solution for the crossing symmetry of the $O(N)$ sum rule \eqref{onsum1}-\eqref{onsum3}, 
this reorganization may be qualitatively different for $N> 0.5$ and $N\ll  0.5$, which may be described as follows.
Suppose we superpose the curves for the sub-leading scaling dimensions on FIG. \ref{fig:kink} and  
 a certain dimension bifurcates to the right (left) as we move to larger (smaller) $\Delta_\phi$; then let us call this $R$-bifurcation ($L$-bifurcation).

Before describing more details of the $O(N)$ spectrum, 
let us mention that the usage of the word ``bifurcation" here does not necessarily mean 
the bifurcation of the common square-root type. We take the following important case to illustrate this.
On the Ising spectrum obtained via the $\mathbb{Z}_2$ sum-rule \eqref{isingsum}, an extensive description is given in Section 3
of Ref. \cite{El-Showk14}, where the recombination is shown to occur \`{a} la Hilbert's infinite hotel.
As both sides of the IR fixed point ($\Delta_\phi ={\Delta_\phi}^*$)
have infinitely many operators, the effective correspondence between the two spectrum can be non-trivial
depending on the versions of the infinite hotels, for which ``$\infty=\infty+1$" and ``$\infty=2\cdot\infty$" are described below.  
In particular, if one temporarily denotes the scalar operators ($\ell=0$) on the right ($\Delta_\phi > {\Delta_\phi}^*$)
by
$\mathcal{E}, \mathcal{E}', \mathcal{E}'', \cdots$ and
those on the left ($\Delta_\phi < {\Delta_\phi}^*$)
by
$\mathcal{E}, \chi, \mathcal{E}', \mathcal{E}''  \cdots$
in ascending order of the scaling dimensions,
the recombination of the spectrum is observed as in the following: 
\begin{align}
\cdots,~ \mathcal{E}'''\overset{d}{\to} \mathcal{E}'',~ \mathcal{E}''\overset{d}{\to} \mathcal{E}',~ \mathcal{E}' \overset{d}{\to} \chi,~ \mathcal{E} \overset{\mathrm{kink}}{\longrightarrow} \mathcal{E}. \qquad (``\infty=\infty+1"),
\label{recombination}
\end{align}
where the symbol ``$\overset{d}{\to}$" stands for a connection of the nearest levels by a sudden descent from the right of ${\Delta_\phi}={\Delta_\phi}^*$ to the left, 
and $\chi$ is a decoupling (null) operator, which only appears numerically on the left side ($\Delta_\phi < {\Delta_\phi}^*$) 
with a small (ideally vanishing) squread OPE coefficient.
Although the nearest levels in this case (e.g. $\mathcal{E}''$ and $\mathcal{E}'$) never touch with each other 
within a finite numerical bootstrap, it is also suggested that the recombination transition becomes shaper
and eventually the nearest levels would be connected in the limit of the infinite number of derivatives ($\mathcal{K}\to \infty$).
In this regard, the connection between both sides for a large, but finite $\mathcal{K}$ would also seem to be, 
\begin{align}
\cdots,~ \mathcal{E}'''\overset{b}{\to} \mathcal{E}'''+\mathcal{E}''\,(+\mathcal{E}'+\chi),~ \mathcal{E}''\overset{b}{\to} \mathcal{E}''+\mathcal{E}'\,(+\chi),~ \mathcal{E}'\overset{b}{\to} \mathcal{E}'+\chi,~ \mathcal{E} \overset{\mathrm{kink}}{\longrightarrow} \mathcal{E}. ~ 
(``\infty=2\cdot \infty"),
\label{bifurcation}
\end{align}
where the levels on the right bifurcate (or possibly multifurcate) to the left as represented by the symbol ``$\overset{b}{\to}$".
Here in the connection $\mathcal{E}''\overset{b}{\to} \mathcal{E}''+\mathcal{E}'$, for instance, the lower branch $\mathcal{E}''\to \mathcal{E}'$ causes a sudden decent by a finite gap $\Delta_{\mathcal{E}''}^{\mathrm{Right}}-\Delta_{\mathcal{E}'}^{\mathrm{Left}}>0$,  
while the upper branch $\mathcal{E}''\to \mathcal{E}''$ (eventually connected 
as $\Delta_{\mathcal{E}''}^{\mathrm{Right}}- \Delta_{\mathcal{E}''}^{\mathrm{Left}}\to -0$
in $\mathcal{K}\to \infty$) does not meet such a large jump if $\mathcal{K}$ is large enough. 
It is also worth to mention that if the decoupling operator $\chi$ on the left indeed disappears 
in the ideal limit $\mathcal{K}\to\infty$, 
the connection is then more like
\begin{align}
\cdots,~ \mathcal{E}'''\to \mathcal{E}''',~ \mathcal{E}''\to \mathcal{E}'',~  \mathcal{E}'\to \mathcal{E}',~ \mathcal{E} \overset{\mathrm{kink}}{\longrightarrow} \mathcal{E},
\label{onetoone}
\end{align}
where all the connections of the levels may be continuous (with no gap left) and may have smoother (or even no) kinks compared with that of the lowest one ($\mathcal{E}$).
In the scalar ($\ell=0$ in $S$-) sector along the unitarity-saturating solutions for the $O(N)$ sum rule \eqref{onsum1}-\eqref{onsum3} around $N=1$, 
we observe a similar recombination as in \eqref{recombination} as expected. 
Besides more common looking bifurcation, below let us simply call these processes in \eqref{bifurcation} e.g. $\mathcal{E}''\overset{b}{\to} \mathcal{E}''+\mathcal{E}'$  a $L$-bifurcation of $\mathcal{E}''$. 

For $N>0.5$, we observe that an $R$-bifurcation of the dimension $\Delta^{(2)}_{A,\ell=1}$ of the sub-leading spin-1 antisymmetric tensor in $A$-sector
\footnote{An analogous R-bifurcation of a spin-1 operator is also observed in the $\mathcal{N}=2$ supersymmetric (SUSY) Ising model \cite{Bobev}, where
the decoupling operator of the lower branch never touches the level of the $J_{ab}^{\mu}$ at $\Delta=2$.
It is also remarkable that 
the $N=0$ model has a twisted $\mathcal{N}=2$ SUSY in 2D \cite{Saleur}, whose origin, 
 the presence of underlying $Osp(2M,2M)$ for any $M$ in $N=0$ \cite{ParisiSourlas}, is actually independent of the space dimension $D$.  
}
(just above the conserved current $J_{ab}^{\mu}$)
and a $L$-bifurcations of the sub-leading dimension $\Delta^{(2)}_S\equiv \Delta_{\mathcal{E}'}\sim 3.8$ 
from $S$-sector 
\footnote{ This $L$-bifurcation of $\Delta^{(2)}_S$ (dimension for $\mathcal{E}'=:E^2:$ with $E=\sum_a \phi_a^2$)
is accompanied by a level crossing of $\Delta^{(2)}_T$ (dimension for $:EF_{ab}:$ with $F_{ab}=\phi_a\phi_b-E/N$)
 and $\Delta^{(3)}_T$ for $N\geqslant 1$.  
In the Ising model,  $\Delta^{(2)}_S$ becomes $\Delta_{\phi^4}\sim 3.8$, which gives the correction to scaling exponent $\omega\sim 0.8$.
In the XY model ($N=2$), we reproduce $\Delta^{(2)}_T\sim 3.65$ \cite{Brezin,CalabresePV2,Echeverri}.
More detailed study of the subleading spectrum is beyond the scope of this work.}
occur simultaneously at  $\Delta_\phi={\Delta_\phi}^*$ of the $O(N)$ CFT (see \eqref{OPE} for the sectors $S$, $T$, and $A$). 
After the $L$-bifurcation, the lower branch of $\Delta^{(2)}_S$ flows into the free value $(\Delta_\phi, \Delta)=(1/2, 2)$ 
(being RG unstable, the free theory at $\Delta_\phi=1/2$ tends to have fast varying subleading dimensions; thus  it is  
numerically subtle to see $\lim_{\Delta_\phi \to 1/2} \Delta^{(2)}_S=2$)
and seem to contribute to larger slopes $\partial\Delta_S/\partial\Delta_\phi$ and $\partial\Delta_T/\partial\Delta_\phi$ in $\Delta_\phi<{\Delta_\phi}^*$ via the level repulsion.

Now for $N\ll 0.5$, 
the $R$-bifurcation does no longer coincide with the $L$-bifurcation; the latter may be observed at much larger $\Delta_\phi$. 
The lower branch after the $R$-bifurcation of the subleading spin-1 dimension flows into the level $\Delta=2$ of the conserved current $J_{ab}^{\mu}$ 
just below in the same $A$-sector.
This isolated $R$-bifurcation and the confluent behavior in $A$-sector should lead to 
the enhanced change of the slope $\partial C_J/\partial\Delta_\phi$ (i.e. the effective asymmetric minimum) at $\Delta_\phi={\Delta_\phi}^*$ of the $O(N)$ CFT 
and the subsequent divergent behavior of $C_J$ in $\Delta_\phi>{\Delta_\phi}^*$, respectively.
The change in the slope $\partial C_T/\partial\Delta_\phi$ seems to be mainly due to the $L$-bifurcation in the $S$-sector;
the $R$-bifurcation in $A$-sector may also change $\partial C_T/\partial\Delta_\phi$, but this effect is rather weak as in the curve for $N=0.1$ in FIG. \ref{fig:cJ}.
The discussion above is obviously not enough to fully describe the level dynamics as one moves along the unitarity bound
and to understand how it may lead to the formation of the kink.
On the other hand, the degeneracy of the levels at $\Delta_\phi={\Delta_\phi}^*$ might play a role in constructing the putative representation theory for 3D CFTs.
Therefore, the bifurcations and the confluent behavior observed here may deserve further studies. 

\subsection{Determination of $\Delta_\phi$ and the fractal dimension in the limit $N\to0$}\label{section:N=0}  
There are practically at least three ways to estimate $\Delta_\phi$ (or equivalently, the anomalous dimension $\eta$) of the $O(N)$ CFT
with some reasonable assumptions for each case: 
\begin{enumerate}
\item Calculate $\Delta_\phi$ that gives the asymmetric minimum of the current central charge $C_J$ (FIG. \ref{fig:cJ}),
\item Calculate $\Delta_\phi$ where the $R$-bifurcation of the subleading spin-1 dimension $\Delta^{(2)}_{A,\ell=1}$ occurs,
\item Locate the kink $(\Delta_\phi, \Delta_T)$ in the unitarity upperbound of $\Delta_T$ (FIG. \ref{fig:kink}).
\end{enumerate}
As discussed in the previous section, the method 1 and method 2 are essentially equivalent and should give consistent estimates.
Within our derivative truncation \eqref{derivativechoice} of $\mathcal{K}\sim 153$, the method 1 is applicable for $N \lesssim 0.4$. 
Although the method 3, which will be used for $N=1$ in Section \ref{section:N=1}, has an advantage of giving simultaneous estimates for $(\Delta_\phi, \Delta_T)$,
it may not be so accurate for $N \lesssim 0.2$ as the smoothing of the kink inevitably occurs (Section \ref{section:kink}).
Taking these into account, using 
the criterions $1$ and $2$ for 
a zoom-up data, one may obtain by the conformal bootstrap
\begin{align}\label{Delta_phi}
\Delta_\phi=0.5145(2) \quad \text{ for } \quad N=0.1,
\end{align}
which amounts, by an ad-hoc linear extrapolation from $\Delta_\phi=0.518151(6)$ for the Ising model ($N=1$) \cite{Simmons-Duffin}, to
$\left.\Delta_\phi\right|_{N=0}=0.5141\pm 0.0002$ for $N=0$ with uncertainty simply copied from \eqref{Delta_phi} as we will not use this later (the notation $\pm x$, instead of $(x)$, is temporally used to indicate uncertainties for the ease of comparison). 
There are various RG estimates for the anomalous dimension $\eta=2\Delta_\phi+2-D$ (see \cite{Pelissetto} for review), which have in general 
relatively larger uncertainties than those for the exponent $\nu^{-1}=D-\Delta_S$. 
Nicely, the estimate for $N=0$ above lies almost at the center of the result $\left.\Delta_\phi\right|_{N=0}=0.5142\pm 0.0013$ obtained from updated and the most accurate RG computation \cite{Guida}.
Note also that MC simulations are rarely able to measure this with an exception of $\left.\Delta_\phi\right|_{N=0}=0.5125\pm 0.0007$ \cite{Liu},
which is slightly smaller than our bootstrap result.

Now, we estimate the optimal unitarity bound for $\Delta_{T}$ at $N=0.1$ with $\Delta_\phi$ in \eqref{Delta_phi} and extrapolate it to $N=0$.
The unitarity upper bound $\Delta_{T}^{*}$ for $\Delta_{T}$ approaches the optimal value from above 
as the number of derivatives $\mathcal{K}$ tends to infinity as shown in Table \ref{tableN0}. 
The downward uncertainty for the last digit ($10^{-5}$) of each $\Delta_{T}^{*}(\mathcal{K})$ due to the choice of the bisection accuracy is shown as a subscript. 
Also note that these last digits may be subject to change by the choice of the cut-off for spins.
As already mentioned, the convergence for this small value of $N$ becomes much slower than that for generic $N$  like  $N=1$.
Note also that ${\Delta_{T}}^{*}(\mathcal{K})$ has a minor variation in addition to the overall tendency to decrease 
(the upper-bound must decrease as the constraints gets stronger).
This is expected since ${\Delta_{T}}^{*}$ depends on the precise choice of the derivatives \eqref{derivativechoice}, 
which has more information than just a one number $\mathcal{K}$.
The uncertainty induced by this variation, however, does not become dominant in the analysis below. 

The estimate for the optimal bound ${\Delta_{T}}^{*}(\infty)$ in Table \ref{tableN0} is obtained by
a phenomenological fit 
using \footnote{It is equivalent to find the intercept at $x=0$ in the linear fit for the data $(x,y)=(1/\mathcal{K}^{p}, {\Delta_{T}}^{*}(\mathcal{K}))$.}
${\Delta_{T}}^{*}(\mathcal{K})={\Delta_{T}}^{*}(\infty) + \frac{\mathrm{const.}}{\mathcal{K}^p}+\mathcal{O}(\frac{1}{\mathcal{K}^{p+1}})$ with $p=2$,
which is presumably better than the raw bound ${\Delta_{T}}^{*}(\mathcal{K}_{\mathrm{max}})$ with $\mathcal{K}_{\mathrm{max}}=153$ obtained by the bisection.
We adopt this value ${\Delta_{T}}^{*}(\infty)=1.2948(36) $ as optimal for $N=0.1$ 
with this conservative error bar, which includes the entire residual $|{\Delta_{T}}^{*}(\mathcal{K}_{\mathrm{max}}) - {\Delta_{T}}^{*}(\infty)|\sim 0.0029$
as well as the smaller uncertainty propagated from \eqref{Delta_phi} along the curve in FIG. \ref{fig:kink}.
Using the above value ${\Delta_{T}}^{*}(\infty)$ for $\Delta_T$ at $N=0.1$ and 
the derivatives $\partial \Delta_T/\partial N$ in \eqref{derivativevalues0}-\eqref{derivativevalues1} in the extrapolation from $N=0.1$,
one may have,  
\begin{align}\label{Delta_T_N=0}
\Delta_T=1.2984(36) \quad \text{ for } \quad N=0,
\end{align}
where the error due to the extrapolation is estimated as the uncertainty in \eqref{derivativevalues0} multiplied by $0.1$ giving $7\times 10^{-4}$, which is negligible compared to the other uncertainties.
We note that the extrapolations from other small values of $N$ would give consistent estimates meaning that
the RG extrapolation by $\partial \Delta_T/\partial N$ is correct and actually avoidable in principle.
With the degeneracy $\Delta_S=\Delta_T$ at $N=0$ understood (Section \ref{section:degeneracy}),
this value \eqref{Delta_T_N=0} is consistent with  
$\Delta_S=1.2999(32)$ ($\nu=0.5882(11)$) from the RG \cite{Guida}
and with $\Delta_S=1.29815(2)$ ($\nu=0.587597(11)$) from the most accurate MC, which is far ahead of other simulations in accuracy \cite{Clisby}.
Using the relation \eqref{d_F}, the symmetric tensor dimension \eqref{Delta_T_N=0} leads to the fractal dimension,
\begin{align}\label{dsaw}
\left.d_F\right|_{N=0}=1.7016(36).
\end{align}
Here let us just mention that this is much larger than the Flory value $d_F=\frac{D+2}{3}=5/3=1.6666\cdots$ \cite{Flory} and should be more precise.
The comparison with the corresponding results for $\Delta_S(=\Delta_T)$ from more modern litertures are just given below \eqref{Delta_T_N=0}.

Now, two remarks are in order.
First, we note that  
using another choice $p=1$ in the fit would lead to 
a larger residual $|{\Delta_{T}}^{*}(\mathcal{K}_{\mathrm{max}}) - {\Delta_{T}}^{*}(\infty)|\sim 0.0085$ resulting in
${\Delta_{T}}=1.293 (9) $ for $N=0$. 
Although this value from $p=1$ is still consistent with the other estimates, one clearly needs to further increase $\mathcal{K}_{\mathrm{max}}$ from $\mathcal{K}_{\mathrm{max}}=153$ in order to obtain a better estimate, which is computationally time consuming.
Although there seems to be no decisive difference between $p=1$ and $p=2$ regarding the quality of the fits,
the fit using the function with the coexisting powers $p=1$ and $p=2$ yields $|{\Delta_{T}}^{*}(\mathcal{K}_{\mathrm{max}}) - {\Delta_{T}}^{*}(\infty)|\sim 0.0024$ giving ${\Delta_{T}}=1.2988(32)$ for $N=0$, which is effectively the same result as \eqref{Delta_T_N=0} obtained with $p=2$ only.
Second, we comment on the subtlety in the analysis for non-integer $N$.  
Before doing so, let us briefly summarise on the three quantities all for $N=0.1$:
${\Delta_{T}}^{*}(\mathcal{K}_{\mathrm{max}})$,
${\Delta_{T}}^{*}(\infty)$,  
$\Delta_{T}$, for which the level of rigor is decreasing in this order.
The unitarity bound ${\Delta_{T}}^{*}(\mathcal{K}_{\mathrm{max}})$
is a rigorous upper bound, albeit not optimal.
The extrapolation to the optimal bound ${\Delta_{T}}^{*}(\infty)$ involves the phenomenologica fit, 
for which the entire residual $|{\Delta_{T}}^{*}(\mathcal{K}_{\mathrm{max}}) - {\Delta_{T}}^{*}(\infty)|$ is included as an error; 
the resulting estimate is not rigorous, but would be called conservative.
Last but not the least, we use ${\Delta_{T}}^{*}(\infty)$ as an estimate for $\Delta_{T}$ in the $O(N)$ CFT. 
This is justified if the unitarity bound is saturated at the $O(N)$ CFT, which is emplically expected at $N\in \mathbb{N}$ as will be done 
for $N=1$ in Section \ref{section:N=1}.
On the other hand, one should not expect an exact saturation for $N=0.1$ as suggested by the unitarity violation in the free $O(N)$ model for non-integer $N$ \cite{Maldacena}. 
We nevertheless expect that $|\Delta_{T} - {\Delta_{T}}^{*}(\infty)|/\Delta_{T}\ll 1$,
which means that the unitarity-saturating solution still passes very close to the location of the $O(N)$ CFT 
as it happens in the bootstrap for the Ising model in non-integer dimensions $D\notin\mathbb{N}$ \cite{El-Showk3}, which is shown to be non-unitary \cite{HogervorstUnitarity}. 
This point is corroborated in Section \ref{section:conclusion}.

\begin{table}
\begin{center}
\begin{tabular}{c||cccccc||c}
$\mathcal{K}$ & 80 & 99 & 117 & 135 & 153 ($\mathcal{K}_{\mathrm{max}})$ & $\infty$ & $\Delta_{\phi}$ \\
\hline
${\Delta_{T}}^{*} (N=0.1)$  & $1.30517_5$ & $1.30096_4$ & $1.30038_4$ & $1.29799_4$ & $1.29770_5$ & 1.2948 & 0.5145\\
${\Delta_{T}}^{*} (N=1)$  & $1.26627_{2}$ & $1.26598_{2}$ & $1.26594_{2}$ & $1.26590_{3}$ & $1.26589_{3}$ & 1.2654 & 0.51815
\end{tabular}
\caption{
The unitarity bound for $\Delta_{T}$ with $N=0.1$ at $\Delta_{\phi}=0.5145$ and with $N=1$ at $\Delta_{\phi}=0.51815$ obtained with various choices $(n_{\text{max}}, m_{\text{max}})$$=$$(7,2)_{80}, (8,2)_{99}, (8,4)_{117}, (8,6)_{135}, (8,8)_{153}$ of the derivatives,
where the number $\mathcal{K}$ of derivatives is given as a subscript.
The data is not ideally smooth as it depends weakly on the precise choice of derivatives \eqref{derivativechoice}, which is not completely specified by $\mathcal{K}$.  For  fixed $\mathcal{K}$, the bound ${\Delta_{T}}^{*}$ 
may be smaller by up to $y \times 10^{-5}$, where $y$ is shown as the subscript of the last digit.  
}
\label{tableN0}
\end{center}
\end{table}

Our purpose here is not a pursuit on the numerical precision, which is presently less than the MC \cite{Clisby}, but is giving a new perspective on how
the conformal bootstrap can be used to determine the fractal dimension. 
Although more sophisticated approaches to the limit $N=0$ deserves further consideration as in Section \ref{section:conclusion},    
the result here may be already encouraging 
enough to let us believe gaining a deeper understanding on the 3D self-avoiding walk based on the conformal invariance is promising.  

\subsection{The end point $N=-2$ and the loop erased random walk}\label{section:N=-2}
The present form of the conformal bootstrap, which depends on the unitarity, can not directly be applied to the $O(-2)$ model,
despite its importance as an endpoint of the continuous $O(N)$ family that would be paired with the spherical model limit $N=\infty$ (Section \ref{section:ONglobal}).
Instead, we compute the pseudo $\epsilon$ expansion ($\tau$-series) for the symmetric tensor dimension $\Delta_T$ in the
3D $O(-2)$ model from the result of the fixed dimension 6-loop RG \cite{Antonenko,CalabresePV}. 
More details can be found in Appendix, where our parallel analysis for the $N$-derivatives of $\Delta_T$ and $\Delta_S$ is performed.
The result is,
\begin{align}\label{tau_N-2}
3-\Delta_{T}=
2-\frac{\tau }{3}-0.0740741 \tau ^2+0.0320229 \tau ^3-0.0226127 \tau ^4+0.0397418 \tau ^5-0.0693066 \tau ^6+O\left(\tau ^7\right)
\end{align}
The $5$- and $6$-loop simple Pad\'{e} analysis parallel to that in Appendix using Table \ref{table:N=-2} yields,   
\begin{align}
\left.d_{F}\right|_{N=-2}=1.614(16)
\label{dlerw}
\end{align}
with which the relatively large uncertainty comes from the oscillating data along the boundary of Table \ref{table:N=-2} (i.e. the direct series $M=0$ and its dual $L=0$) and is
estimated as a root-mean-square deviation.
As expected, the estimate \eqref{dlerw} is
consistent with the fractal dimensions in the loop-erased random walk (LERW) obtained in a number of works including
$d_{\text{LERW}}=1.614(11)$ by the functional RG \cite{Fedorenko} and by various simulations:
$d_{\text{LERW}}=1.623(11)$ \cite{Guttmann},  
$1.6183(4)$ \cite{Agrawal},  
$1.6236(4)$ \cite{Grassberger},  
$1.62400(5)$ \cite{Wilson}.
As a remark on the analysis of the Pad\'{e} table, it may be possible to note that field theories tend to prefer a slightly smaller central values 
compared with the numerical predictions.
In our case, omissions of the four boundry data (indicated by $^*$ in Table \ref{table:N=-2}) would lead to 
$\left.d_{F}\right|_{N=-2}=1.6162(23)$, which would agree with the functional RG \cite{Fedorenko} and older simulations \cite{Guttmann,Agrawal}, but would be definitely smaller than the most recent numerical results \cite{Grassberger,Wilson}. 
It would be interesting to improve the present conformal bootstrap so as to analyse $N<0$
across the severe unitarity wall, to obtain a better estimate on $\left.d_{F}\right|_{N=-2}$, and 
the dimension for sub-leading operators that is responsible for the correction to scaling.
Such study may contribute to deeper understanding on the LERW from the conformal invariance in 3D.

\begin{table}
\begin{center}
\begin{tabular}{c|ccccccc}
$M\backslash L$   &0&1&2&3&4&5&6 \\
\hline
0& 2 & 1.66667 & 1.59259 & 1.62462 & 1.60200 & $^*$1.64174 & $^*$1.57244 \\
1& 1.71429 & 1.57143 & 1.61495 & 1.61136 & 1.61642 & 1.61649 & \text{--} \\
2& 1.62406 & 1.62277 & 1.61106 & 1.61385 & $\underline{1.61649}_{[108]}$ & \text{--} & \text{--} \\
3& 1.62279 & 1.62418 & 1.61643 & 1.61622 & \text{--} & \text{--} & \text{--} \\
4& 1.60818 & 1.61698 & $\underline{1.6162}_{[14]}$ & \text{--} & \text{--} & \text{--} & \text{--} \\
5& $^*$$\underline{1.63063}_{[2.7]}$ & 1.61633 & \text{--} & \text{--} & \text{--} & \text{--} & \text{--} \\
6& $^*$1.59218 & \text{--} & \text{--} & \text{--} & \text{--} & \text{--} & \text{--} \\
\end{tabular}
\caption{\footnotesize
The Pad\'{e} table for $\left.d_{F}\right|_{N=-2}=3-\Delta_T$. The positive real pole closest to $1$ is shown in the bracket.}
\label{table:N=-2}
\end{center}
\end{table}

\subsection{Fractal dimension of the high-temperature graphs in the 3D Ising model}\label{section:N=1}
The fractal dimension of the critical excitation in the $O(1)$ model 
is most straightforwardly and rigorously accessible by the conformal bootstrap since the bisection for $\Delta_T$ 
yields a clear kink along the unitarity bound just as in the conventional cases $N \geqslant 2$ \cite{Kos14}
and since the $O(1)$ model is unitary so that the $O(1)$ IR fixed point 
may be expected to saturate the optimal unitarity bound as in the Ising case.
By a brief inspection of the zoom-up of FIG. \ref{fig:kink}, one obtains  $\Delta_T\sim 1.266$ and $\Delta_\phi\sim 0.5181$,
for which the latter is consistent with the known result for the Ising spin operator $\Delta_\phi=0.518151(6)$ \cite{Simmons-Duffin}.
It is also empirically interesting, though being not rigorous at all, that
if we adopt an ad-hoc criterion that the kink is located around 
$\Delta_\phi$ where the bisection takes longer than a certain period ($\sim 2$ weeks, for instance), we obtain another precise estimate
$\Delta_\phi=0.518149(6)$ from the $O(N)$ sum rule \eqref{onsum1}-\eqref{onsum3} with $N=1$; 
near the kink, the optimization takes indeed longer time than the other generic points, while
the convergence to the optimal bound with respect to the number of derivatives $\mathcal{K}$ becomes much faster. 

\begin{figure}
\includegraphics[width=16cm]{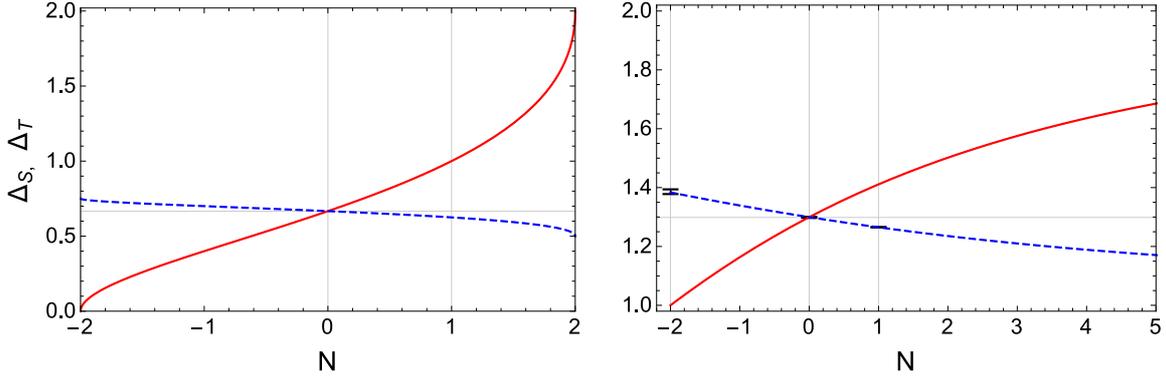}   
	\caption{ \footnotesize
The scaling dimensions $\Delta_S$ (solid) and $\Delta_T=D-d_F$ (dashed) as a function of $N$ in 2D (left: exact) and in 3D (right: $[4/1]$-Pad\'{e}).
Only for $N=-2$ (LERW) in 3D, the error bar is visible where $\Delta_T$ is obtained by the RG.
For $N=0$ (SAW) and $N=1$ (Ising high-$T$ graph),  $\Delta_T$ are obtained by the 3D conformal bootstrap where the error bars are less than the line width.  
               \label{fig:NDelta} } 
\end{figure}

The unitarity upper bound ${\Delta_T}^{*}$  for $N=1$ at $\Delta_\phi=0.51815$
for various choices of the derivatives, represented by its number of the components $\mathcal{K}$, are shown in Table \ref{tableN0}.
This leads via \eqref{d_F} to an estimate for the
fractal dimension for the high-temperature graphs in the 3D Ising model,
\begin{align}
\left.d_{F}\right|_{N=1}=1.7346(5),
\label{dising}
\end{align}
which agrees well with $d_F=1.7349(65)$  from the 3D Ising plaquette-update MC simulation \cite{Winter} and
with the worm algorithm simulation $d_F=1.734(4)$ \cite{Liu}.
Our results by the conformal bootstrap ($N=0$ \eqref{dsaw} and $N=1$ \eqref{dising})  and by the RG ($N=-2$ \eqref{dlerw}) are shown in FIG. \ref{fig:NDelta}.
The related estimates using \eqref{d_F} for the XY model ($N=2$) is given in a footnote in Section \ref{section:conclusion}.

Since the $O(1)$ model is unitary, the rigorous upper-bound ${\Delta_{T}}^{*}(\mathcal{K}_{\mathrm{max}})$
applies to the $O(1)$ CFT, which yields $d_F=3-{\Delta_{T}}^{*}(\mathcal{K}_{\mathrm{max}})>1.7341$.
In contrast to the case with $N=0.1$ in Section \ref{section:N=0},
it would be natural to assume that the $O(1)$ CFT saturates the optimal bound.
Note also that the uncertainty $6\times 10^{-6}$ in $\Delta_{\phi}$, as being multiplied by $(\partial \Delta_T/\partial \Delta_{\phi})|_{\Delta_{\phi}=0.51815}$,
may induce the error only less than $10^{-4}$ in $d_F$.  
Then the source of uncertainty is virtually restricted to the extrapolation of ${\Delta_{T}}^{*}$ 
 from $\mathcal{K}=\mathcal{K}_{\mathrm{max}}$ to $\mathcal{K}=\infty$.
This uncertainty 
$|{\Delta_{T}}^{*}(\mathcal{K}_{\mathrm{max}}) - {\Delta_{T}}^{*}(\infty)|\sim 5\times 10^{-4}$
as shown in \eqref{dising} is evaluated from the fit using
${\Delta_{T}}^{*}(\mathcal{K})={\Delta_{T}}^{*}(\infty) + \frac{\mathrm{const.}}{\mathcal{K}^p}+\mathcal{O}(\frac{1}{\mathcal{K}^{p+1}})$
with $p=1$. 
We keep this conservative estimate, while the uncertainty is roughly halved if one uses the fit with $p=2$ 
or becomes even smaller if one uses an exponential fit, for which the quality of the latter seems decent for this particular case.  
It would be very useful if one had a general theory for the scaling of the residual with respect to the optimal bound as a function of $\mathcal{K}$.
In any case, the conformal bootstrap here for this particular fractal dimension seems to give an order of magnitude more precise result  
compared with the MC simulations \cite{Winter,Liu}.

\subsection{The tricrtical Ising fixed point and the $\mathcal{N}=1$ SCFT}\label{section:SUSY}
In 2D, it is well-known that the CFT for the trictitical Ising model ($c=7/10$) is known to be 
the first member of the series of the minimal $\mathcal{N}=1$ superconformal CFTs (SCFTs) \cite{Friedan}.
In condensed matter, the $\mathcal{N}=1$ SCFT in 3D is propsed to describe boundary excitations in the topological superconductor \cite{Grover}.
The action of the one-component Gross-Neveu-Yukawa model in $D$-dimensions is given by
\begin{align}\label{action}
S= \int \,d^Dx\, \,  [  \bar{\psi} \displaystyle{\not}\partial \psi
+ g\phi \bar{\psi}\psi  
+\frac12 (\partial \phi)^2   +\frac{r}{2} \phi^2   +u\phi^4], 
\end{align}
where the Yukawa coupling mixes the bosons $\phi$ and the fermion $\psi$. At the $\mathcal{N}=1$ supersymmetric 
IR fixed point,
the interaction is described in terms of the superpotential $W=\Sigma^3$ and a real-supermultiplet  
$\Sigma = \phi +\theta \psi +\theta^2 \phi^2$ \cite{Bashkirov}, 
where 
$\theta$ is a fermionic coordinate (in superspace) of dimension $-1/2$ regardless of the space dimension $D$ and
$\theta^2=\epsilon^{\alpha\beta}\theta_{\alpha}\theta_{\beta}$ ($\alpha, \beta=1, 2$).
Thus it is pointed out that the intersection of the following extra constraint
\begin{align}\label{SUSYrelation}
\Delta_{\phi^2}^{\mathcal{N}=1}=\Delta_{\phi}^{\mathcal{N}=1}+1
\end{align}
with the unitarity bound curve $\Delta_{\phi^2}=\Delta_{\phi^2}(\Delta_{\phi})$  
may be used to locate the SCFT \cite{Bashkirov}. 

In contrast to the branch of this curve in the left of the kink, 
the right branch in the relevant region, which intersects with \eqref{SUSYrelation}, is observed to be almost linear both in 2D and 3D
\cite{El-Showk14}. Let us then define the coefficients $a$ and $b$ of the linear fit in 3D:
\begin{align}\label{unitarityboundfit}
\Delta_{\phi^2}\sim a \left(\Delta_{\phi}-\frac{1}{2}\right) +b.
\end{align}
For each choice of the set of the derivatives $(n_{\text{max}}, m_{\text{max}})=(2k, 1)$, ($k=4,\,5,\cdots, 10$) 
with the cut-off $\mathcal{K}$ for the number of derivatives given in Section \ref{section:bootstrap} and 
the spin cut-off at $L=50$ for $k=9,10$ and at $L=40$ for otherwise, 
the linear fit is performed for the right branch of the 
the unitarity upper-bound of $\Delta_{\phi^2}$
in the range $\Delta_\phi \in [0.54, 0.59]$ obtained through the $\mathbb{Z}_2$ sum rule.
Table \ref{SUSYtable} shows the position of the intersection $\Delta_{\phi}^{\text{cross}}$ 
as well as the coefficients of the fit
\footnote{It would be interesting to study this slope $a$ as a function of $D<4$, though it is beyond the scope here (see \cite{El-Showk3} for related figures). One ad-hoc interpolation between the two values $a\sim 3$ ($D=3$) and $a=8/3$ ($D=2$) is $a=2(6-D)/(5-D)$ inspired by the approximate coincidence of $a$ with the critical dimensions where $\phi^6$ and $\phi^8$ becomes RG marginal.}
 with respect to $\mathcal{K}$.
Although the convergence of the data may be improved by choosing 
the cut-offs (and the range of the fit) which consumes much more computation time, this would not seriously change the qualitative argument below.  
Table \ref{SUSYtable} may lead to an estimate
\begin{align}\label{SUSYphi}
\Delta_{\phi}^{\mathcal{N}=1}\sim 0.574, \quad  \Delta_{\phi^2}^{\mathcal{N}=1}=\Delta_{\phi}^{\mathcal{N}=1}+1,
\end{align}
which is slightly larger than the one-loop RG result $1/2+1/14=0.571\cdots$ \cite{Grover} and satisfies
the rough lower bound $\Delta^{\mathcal{N}=1}_{\phi}>0.565$ \cite{Bashkirov}
\footnote{The same approximate value $\Delta_{\phi}=0.565$ has recently been reproduced by the fermion bootstrap \cite{Iliesiu}.}
.
	
\begin{table}
\begin{center}
\begin{tabular}{c|cccccccc}
$\mathcal{K}$ &56&90 & 132 & 182 & 240  & 306 & 380 \\ 
\hline 
$\Delta_{\phi}^{\text{cross}}$ &0.5651(1) &0.5689(2) &  0.5705(2) & 0.5713(2)& 0.5722(4) &0.5731(6) & 0.5737(5) \\
$a$& 3.2454(16)&3.1209(25)&3.0679(26)&3.0443(27)&3.0323(51)&3.0235(80)&2.9955(64)\\
$b$& 1.3538(1)&1.3537(2)&1.3541(2)&1.3542(2)&1.3532(4)&1.3520(6)&1.3528(5)
\end{tabular}
\caption{\footnotesize
The cut-off $\mathcal{K}$ of derivatives  and 
the corresponding position $\Delta_{\phi}^{\text{cross}}$ of the intersection between the SUSY relation \eqref{SUSYrelation}
and the linear fit \eqref{unitarityboundfit} for the unitarity bound for $\Delta_{\phi^2}$ as well as the coefficients $a$ and $b$.}
\label{SUSYtable}
\end{center}
\end{table}

The geometric exponent for the $\mathcal{N}=1$ fixed point may be obtained by the conformal bootstrap giving the symemtric tensor dimension
$\Delta_T\sim 1.43$ in the $O(1)$ model
at $\Delta_\phi=\Delta_{\phi}^{\mathcal{N}=1}$. 
The derivation of \eqref{d_F} does not seem to depend on 
whether a fixed point may allow the description by the supersymmetry or not, 
although more detailed analysis would be useful.     
By assuming \eqref{d_F} holds here also, one would obtain,  
\begin{align}\label{SUSYfractal}
\left.d_F\right|_{SUSY}\sim 1.57.
\end{align}

In order to find the fractal object with the dimension \eqref{SUSYfractal} in a lattice model, 
one natural candidate would be the Kitaev model 
augmented by a local Ising exchange interaction at finite temperature, in which the non-supersymmetric fractal dimension 
has already been realized by the magnetic flux loops \cite{Kamiya}.
In particular, the critical temperature along the paramagnetic to quantum-spin-liquid transition was determined by 
identifying the magnetic flux loop in this extended Kitaev model (an effective $\mathbb{Z}_2$ gauge system) with 
the 3D Ising high-temperature graphs 
and by using the knowledge \cite{Winter} of the fractal dimension for the latter, for which our independent estimate is in \eqref{dising}.
Similarly, it would be interesting if the fractal dimension \eqref{SUSYfractal} is realized by the flux loop at some point 
in the phase diagram (or its suitable extension) and can be used to locate the emergent $\mathcal{N}=1$ fixed point by simulation. Indeed, the phase diagram 
is already very rich and studied for understanding
the effect of thermal agitations on the topological order, and more specifically, the thermal fractionalization of the quantum spins into Majorana fermions.
It is also worthwhile to note that if the $\mathcal{N}=1$ SCFT is realized in the vicinity of  the ``tricritical" point 
\footnote{Numerical accuracy in the Monte-Carlo simulation \cite{Kamiya} has still to be improved to measure the exponents at 
this particular tricritical point. We thank the author, Y. Kato for useful discussions and clarification on the related work \cite{Kato}.}
in \cite{Kamiya},
the corresponding RG fixed point may be different from that of the 
3D Ising tricritical point, which is widely believed to be described by the mean field exponents 
in view of the RG argument including $\phi^6$ interaction
\cite{Riedel}.

Another interesting model which would be relevant to the $\mathcal{N}=1$ supersymmetry is the 
Blume-Capel model \cite{Blume}.
In 2D, the tricritical point of this model belongs to the tricritical Ising universality class, 
which can be identified with the celebrated
 $\mathcal{N}=1$ fixed point \cite{Friedan}. 
In the latter, the set of scaling dimensions which is obtained as an intersection 
between the relation \eqref{SUSYrelation} and the unitarity bound (which is expected to be saturated by the analytic solution \cite{Liendo} to the crossing symmetry along 
$\Delta_{\phi^2}=\frac{8}{3}\Delta_{\phi}+\frac{2}{3}$)
is actually $(\Delta_\varepsilon,\Delta_{\varepsilon'})=(1/5,6/5)$, where the energy operator $\varepsilon$ and subleading energy operator $\varepsilon'$ 
both in the $\mathbb{Z}_2$-even sector
together form one operator in the Neveu-Schwarz sector (while the spin operator, which is $\mathbb{Z}_2$-odd, independently belongs to the Ramond sector with dimension $\Delta_\sigma=3/40$ \cite{Friedan}). 
Thus the correspondence $(\Delta_\phi, \Delta_{\phi^2})\to (\Delta_\sigma, \Delta_\varepsilon)$
in the Ising model should be replaced by $(\Delta_{\phi}^{\mathcal{N}=1}, \Delta_{\phi^2}^{\mathcal{N}=1})\to (\Delta_\varepsilon, \Delta_{\varepsilon'})$ in the 
$\mathcal{N}=1$ model in 2D.
Regarding the crossing symmetry, this can happen in 2D because $\sigma$ in the Ising model and $\varepsilon$ in the 
$\mathcal{N}=1$ model have the same fusion rule since they belong to the same position $(r,s)=(1,2)$ in the Kac table of the Virasoro representation. 
If we use the same identification $\Delta_{\phi}^{\mathcal{N}=1}\to \Delta_\varepsilon$ as in 2D, this gives the exponents
in the thermal sector
\footnote{Another identification $\Delta_{\phi^2}^{\mathcal{N}=1}\to \Delta_\varepsilon$ yields $\nu \sim 0.701$,
which seem to agree with four estimates for $\nu$ ($\sim 0.71$) by the functional RG \cite{Hellwig} for the $\mathcal{N}=1$ UV Lagrangian. 
Note that the latter $\nu$ is not meant for a physical realization (e.g. the Blume-Capel model) and is just an 
indication that $\Delta_{\phi^2}^{\mathcal{N}=1}$ computed from the mass renormalization may agree with the conformal bootstrap. In the 2D $\mathcal{N}=1$ fixed point, for instance, the observed value $2-\alpha=d\nu=10/9\sim1.11$ \cite{Burkhardt} follows from 
$\Delta_{\phi}^{\mathcal{N}=1}=1/5$, but not from $\Delta_{\phi^2}^{\mathcal{N}=1}$.
}
\begin{align}\label{SUSYalphanu}
2-\alpha\sim 1.236, \quad \nu\sim 0.412,  
\end{align}
which are in a reasonable agreement with the exponent $2-\alpha=1.213$ found by the variational RG method 
in the phase diagram of the 3D Blume-Capel model \cite{Burkhardt}.
As noticed in \cite{Burkhardt}, the latter value, which was aimed at the 3D tricritical exponent,
deviates considerably from the standard Ising tricritical exponent, which is believed to be the mean field value $2-\alpha=3/2$ \cite{Riedel} 
as already mentioned,
while their results agree impressively 
with the 2D exact results for both the tricritical and the critical point, and with the modern estimate for the 3D critical point.
This seems to leave some possibility that their method actually detects an additional $\mathcal{N}=1$ fixed point 
with the exponent \eqref{SUSYalphanu} in the Blume-Capel model
other than the possibility that their value quoted above is simply a very poor estimate for the 3D Ising tricritical point. 
In any case, in view of the RG,
it is likely that it is only in $D=2$ that the Ising tricritical point and the $\mathcal{N}=1$ fixed point can be identified with each other. 
It is interesting to study how these two fixed points would deviate from each other in $D=2+\epsilon$ 
\cite{Gracey,Hikami92}
and evolve 
all the way to the different universality classes in $D=3$ by the conformal bootstrap.

\section{Conclusion}\label{section:conclusion}
We take the simplest bootstrap approach based only on the crossing symmetry of the four-point function 
$\langle \phi_a\phi_b\phi_c\phi_d \rangle$ of the fundamental fields for the one-parameter-family of the 3D $O(N)$ model 
with a special focus on the fractal dimension $d_F$ for the range $0\leqslant N \leqslant 1$. 
Besides the property of being exactly at the severe unitarity wall (Section \ref{section:kink}),
the limit $N\to 0$ may be characterized by  the degeneracy of the two operator dimensions $\Delta_S$ and $\Delta_T$ in any space dimensionality $D$.
Accordingly, a more elaborate approach to such limits would need to deal with the possible logarithms that could appear in the four-point functions 
\cite{CardyLog,Vasseur,HogervorstLog}.
Also if one tries to perform the mixed correlator bootstrap \cite{Kos15,Kos16} including the energy fields $\varepsilon$, 
for which the logarithms appear at the level of two point functions,
it would be inevitable to face with these logarithms. 
In that case, the smooth continuation by the conformal bootstrap to the 2D problem from $D=2+\epsilon$ would be also interesting 
since the correlation function including $\varepsilon$ in the 2D $O(N)$ model can be dealt with 
both in the integral-representation \cite{Dotsenko} and the differential equation  
using the degenerate representation \cite{Belavin} at the integer level $3$ in the Virasoro algebra.
This is a subject of further research.

The issue of unitarity violation is certainly important and nontrivial, and it is currently not obvious to what extent it affects the estimates
in Section \ref{section:N=0} for non-integer $N$ obtained using the assumption of the unitarity, which is actually weakly violated 
as outlined below.
Nevertheless, there seems to be various possible improvements for the study of the bootstrap for non-unitary systems 
(e.g. the determinant method \cite{Gliozzi} or the extremal function flow method \cite{PaulosGoFlow})
suggested from the present work. 
To elaborate on some of these, let us first recall that 
the positivity is violated in 
the free $O(N)$ models at any non-integer value of $N$,
which can be shown by computing the norms 
\footnote{
These norms are considered
in the AdS$_4$/CFT$_3$ holography for large $N$.
We thank Tassos Petkou for pointing out that there would be an interesting physics also for small $N$, 
where the bulk degrees of freedom in this higher-spin holography may be related with 
a polymer-like product of the singletons \cite{Leigh} in the replica limit.
}
for a certain class of composite operators \cite{Maldacena}.
Although the above argument only applies to the free model, 
the recent study on the unitarity violation in the single scalar $\phi^4$-theory ($N=1$ is fixed) in the fractional dimensions $D=4-\epsilon$ points to a quatitatively similar behavior, both for the free (UV) theory and for the interacting (IR) theory,  
that the positivity violation indeed occurs, but generically only at the operator with very high scaling dimensions $\Delta \gg  D$
\cite{HogervorstUnitarity},
which would explain why the bootstrap for the Ising model in non-integer dimensions works decently \cite{El-Showk3}. 
Our exact computation for the squared OPE coefficients   
in the 2D $O(N)$ model for non-integer $N$ 
also shows a similar behavior for the positivity violation.
In view of these analysis,
we conclude that the set of the scaling dimensions $(\Delta_{\phi}, \Delta_{T})$ of the IR $O(N)$ FP for non-integer $N$
should not saturate the unitarity upper bound $\Delta_{T}\leqslant \Delta_{T}^{*}(\Delta_{\phi})$ in general,
but hopefully it lies in the vicinity of the bound (it could appear on either side of the bound), 
for $N$ not too close to $N=0$, at which the squared OPE coefficient for the low-lying operator, namely, 
the stress-energy tensor $T^{\mu\nu}$ ($\Delta=D$) changes its sign due to the pole as discussed in Section \ref{section:convergence}.
For $N\sim 0$, on the other hand,   if the family of the unitarity saturating solution obtained here along $\Delta_{T}=\Delta_{T}^{*}(\Delta_{\phi})$
passes indeed nearby the IR $O(N)$ FP,
an inclusion of one more spin-$1$ operator just above the conserved current $J_{[ab]}^{\mu}$
in the truncated spectrum in the determinant method \cite{Gliozzi} 
should qualitatively improve the behavior of the non-unitary solution according to the mechanism in Section \ref{section:slopechange} 
for the emergence of the kink in the current central charge $C_J$.
It would be nice to quantify the effect of the unitarity violation in more detail and 
clarify how (via the flow method \cite{PaulosGoFlow} for instance) the putative non-unitary IR $O(N)$ FP and the unitary solution here could be connected with each other.

In the long run, it would be interesting to generalize the ideas in the SLE so as to describe the critical geometry embedded in 3D 
though the task would severely face to our limited understanding on the 3D geometry
since the SLE is intrinsically based on the Riemann mapping (uniformization) theorem on the 2D conformal map.
In this respect, the 3D $O(N)$ model with $-2\leqslant N \leqslant \infty$ offers a natural one-parameter-family of the loop ensembles (\cite{Liu} for an extensive simulations), which is likely to have a conformally invariant measure (\cite{Kennedy} for a simulation at $N=0$). 
We have been able to focus on the region $0 \leqslant N \leqslant 1$, which have the most interesting cases 
\footnote{The same idea applies to the case $N >1$. For some $N\geqslant 2$, $\Delta_T$ has been determined from the conformal bootstrap \cite{Kos14}. For instance, we will have $d_F=1.76437(108)$ for the high-temperature graphs in the XY model ($N=2$),
which agrees with the simulations $d_F=1.7626(66)$ \cite{Winter}, $1.7655(20)$ \cite{Prokofev}, and $1.765(3)$ \cite{Liu}.  
}
as the two boundary points (namely, the self-avoiding-walks in the $N\to 0$ model and  the high-temperature graphs in 3D Ising model in the $N=1$ model)  
and to estimate the fractal dimension $d_F=3-\Delta_T$ by the conformal bootstrap using the unitarity conditions.
Although the positivity \eqref{positivity} prevents us from applying the conformal bootstrap across the severe unitarity wall at $N=0$,
we compute the fractal dimension for $N=-2$ by the 6-loop RG giving $d_F\sim 1.614$, which would be encouraging to conjecture that 
the 3D $N=-2$ model may also describe the loop erased random walk as in 2D although some elaborate 
operator correspondence may be necessary in view of the logarithmic corrections \cite{Fedorenko}.
It is of interest to see if some generalization of the Beffara's theorem \eqref{Beffara} exists in 3D and 
if the loop ensemble in the 3D $O(N)$ model with $-2\leqslant N \leqslant \infty$ 
of the fractal dimension $1.614\lesssim d_F \leqslant 2$ (parallel to $5/4\leqslant d_F \leqslant 3/2$ in 2D)
has a natural parametrization in terms of the inverse-trigonometric \eqref{kappa}, or other transcendental function of $N$, as $\kappa$ of the 2D SLE$_\kappa$.

\begin{acknowledgements}
The work of H.~S. is supported by JSPS KAKENHI Grant-in-Aid 15K13540.
S.~H. and H.~S. are supported by JSPS KAKENHI Grant-in-Aid 16K05491. 
We thank \'{E}douard Br\'{e}zin, Ferdinando Gliozzi, Yasuyuki Kato, Yoshitomo Kamiya, Shinsuke Kawai,
Andreas L\"{a}uchli, Jonathan Miller, Yu Nakayama, Tassos Petkou, Hirotaka Sugawara, and Slava Rychkov for valuable discussions and correspondences. We thank useful comments from Nikolay Bobev, Giacomo Gori, Matthijs Hogervorst, Jesper Jacobsen, Tomoki Ohtsuki, Miguel Paulos, Marco Serone, Andrea Trombettoni, and Alessandro Vichi. 
We thank the Galileo Galilei Institute for Theoretical Physics for the
hospitality and the INFN for partial support during the completion of this work.
\end{acknowledgements}


\appendix
\section{Computations on the $N$-derivatives by the fixed dimension RG}
We use the fixed dimension RG ($D=3$) augmented by the pseudo-$\epsilon$ series \cite{LeGuillou} to circumvent the accumulation of the intermediate systematic error
due to the determination of the coupling $g^*$ at the non-trivial fixed point. 
The beta function is generalized by a new parameter $\tau$ such that at $\tau=1$ it reduces to the original beta function $\beta(g)$ in 3D:
\begin{align}
\beta(g,\tau)= - \tau g + \beta_2(g),
\label{betagtau}
\end{align}
where $\beta_2(g)\equiv  \beta(g)+g$ starts at order $g^2$ with a positive coefficient of order $1$.
Then the critical exponents can be expanded in $\tau$ by eliminating $g$ by using $g=g^*(\tau)$ which solves $\beta(g^*,\tau)=0$.
We compute the $\tau$-series for the derivatives of $\Delta_S$ and $\Delta_T$ at the degeneration point ($N=0$) based on the
six-loop 3D RG results for $\beta(g)$, $\eta$, $\gamma^{-1}$ \cite{Antonenko} and $\eta_T=\eta+\phi_2/\nu-2$ \cite{CalabresePV}.  
The results are,
\footnote{In the process, this computation naturally reproduces the $\tau$-series for $y_2=3-\Delta_T$ in \cite{CalabreseP}.}
\begin{align}
\left. \frac{\partial}{\partial N}\right|_{N=0}\hspace{-5mm}\Delta_S=&\frac{3}{32}\tau+0.0241609 \tau ^2+0.0036762 \tau ^3+0.0021794 \tau ^4+0.0004817 \tau ^5-0.0019009 \tau ^6+\mathcal{O}\left(\tau ^7\right) \nonumber\\
=&\frac{3}{32} \epsilon+0.0361328 \epsilon ^2-0.0198967 \epsilon ^3+0.0380668 \epsilon ^4-0.0611648 \epsilon^5+\mathcal{O}\left(\epsilon^6\right),\\ 
\left. \frac{\partial}{\partial N}\right|_{N=0}\hspace{-5mm}\Delta_T=&-\frac{1 }{32}\tau-0.0099826 \tau ^2 +0.0049121 \tau ^3-0.0029128 \tau ^4+0.0066431 \tau ^5-0.0101052 \tau ^6+\mathcal{O}\left(\tau ^7\right) \nonumber\\
=&-\frac{1}{32} \epsilon -0.0146484\epsilon^2+0.032921\epsilon^3+0.028875\epsilon^4+
\mathcal{O}\left(\epsilon^5\right),
\end{align}
where we also show the derivatives computed from $\epsilon=4-d$ expansion up to known orders \cite{Kleinert,Kirkham} just for comparison.
Since both $\tau$-series do not show strong asymptotic behaviors with factorial growth of coefficients up to the orders presented,
even the naive direct summation of the series would be of some use; in particular, it is clearly better than the direct sum of the $\epsilon$-expansion.

\begin{table}
\begin{center}
\begin{tabular}{c|cccccc}
$M\backslash L$   &1&2&3&4&5&6 \\
\hline
0 &0.09375 & 0.117911 & 0.121587 & 0.123766 & 0.124248 & 0.122347 \\
1 &0.126299 & 0.122247 & 0.126939 & $\underline{0.124385}_{[4.5]}$ & 0.123864 & -- \\
2 &0.121834 & 0.123695 & $\underline{0.124545}_{[3.0]}$ & $\underline{0.135602}_{[1.3]^*}$ & -- & -- \\
3 &0.125111 & $\underline{0.124448}_{[3.4]}$ & $\underline{0.12388}_{[4.2]}$ & -- & -- & -- \\
4 &0.124285 & $\underline{0.125378}_{[0.2]}$ & -- & -- & -- & -- \\
5 &0.121 & -- & -- & -- & -- & -- \\
\end{tabular}
\caption{\footnotesize
The Pad\'{e} table for the derivative $\partial\Delta_S(0)/\partial N$. The positive real pole closest to $1$ is shown in the bracket.}
\label{table1}
\end{center}
\end{table}
\begin{table}
\begin{center}
\begin{tabular}{c|cccccc}
$M\backslash L$   &1&2&3&4&5&6 \\
\hline
0& -0.03125 & -0.0412326 & -0.0363206 & -0.0392334 & -0.0325903 & -0.0426955 \\
1& -0.0459184 & -0.0379405 & -0.0381491 & -0.0372085 & -0.0365985 & -- \\
2& -0.0332523 & -0.0381572 & -0.0379608 & $\underline{-0.036274}_{[2.7]}$ & -- & -- \\
3& -0.0437948 & -0.0366876 & -0.0367275 & -- & -- & -- \\
4& -0.029235 & $\underline{-0.0367284}_{[45]}$ & -- & -- & -- & -- \\
5& $\underline{-0.0619539}_{[1.2]^*}$ & -- & -- & -- & -- & -- \\
\end{tabular}
\caption{\footnotesize
The Pad\'{e} table for $\partial\Delta_T(0)/\partial N $. The positive real pole closest to $1$ is shown in the bracket.}
\label{table2}
\end{center}
\end{table}
A simple Pad\'{e} analysis, however, may improve the stability of analysis as usual.
This can be irrustrated as follows.
We show the values from the Pad\'{e} approximants $[M/L]$ for these derivatives in Table \ref{table1} and Table \ref{table2},  respectively.
The positive real poles closest to $\tau=1$ are shown in brackets for the six-loops (anti-diagonals $L+M=6$) and five-loops ($L+M=5$) order approximants. 
For each derivative, the data occuring with a pole in $[0.5,1.5]$ (indicated by $^*$) is omitted since it is rather close to $\tau=1$, where the series is to be evaluated. 
As a simple estimate, we take the average of the six and five-loops and the maximum deviation as an error. This gives the value quoted in 
\eqref{derivativevalues0} and \eqref{derivativevalues1} in the text.

\end{document}